\documentclass{article}
\usepackage[utf8]{inputenc}
\usepackage{graphicx}
\usepackage{amssymb}
\usepackage{url,lineno,microtype}
\usepackage{booktabs}
\usepackage{hyperref}
\usepackage{subcaption}
\usepackage{authblk}

\title{Multi-Modal Data Collection for Measuring Health, Behavior, and Living Environment of Large-Scale Participant Cohorts: Conceptual Framework and Findings from Deployments}
\author[1]{Congyu Wu\footnote{Correspondence: congyu.wu@austin.utexas.edu}}
\author[2]{Hagen Fritz}
\author[2]{Zoltan Nagy}
\author[2]{Juan P. Maestre}
\author[3]{Edison Thomaz}
\author[3]{Christine Julien}
\author[4]{Darla M. Castelli}
\author[1]{Kaya de Barbaro}
\author[6]{Gabriella M. Harari}
\author[5]{R. Cameron Craddock}
\author[2]{Kerry A. Kinney}
\author[1,7]{Samuel D. Gosling}
\author[1]{David M. Schnyer}
\affil[1]{Department of Psychology, University of Texas at Austin}
\affil[2]{Department of Civil, Environmental, and Architectural Engineering, University of Texas at Austin}
\affil[3]{Department of Electrical and Computer Engineering, University of Texas at Austin}
\affil[4]{Department of Kinesiology and Health Education, University of Texas at Austin}
\affil[5]{Department of Diagnostic Medicine, University of Texas at Austin}
\affil[6]{Department of Communication, Stanford University}
\affil[7]{Melbourne School of Psychological Sciences, University of Melbourne}

\date{October 2020}

\begin{document}

\maketitle

\begin{abstract}

As mobile technologies become ever more sensor-rich, portable, and ubiquitous, data captured by smart devices are lending rich insights into users' daily lives with unprecedented comprehensiveness, unobtrusiveness, and ecological validity. A number of human-subject studies have been conducted in the past decade to examine the use of mobile sensing to uncover individual behavioral patterns and health outcomes. While understanding health and behavior is the focus for most of these studies, we find that minimal attention has been placed on measuring personal environments, especially together with other human-centric data modalities. Moreover, the participant cohort size in most existing studies falls well below a few hundred, leaving questions open about the reliability of findings on the relations between mobile sensing signals and human outcomes. To address these limitations, we developed a home environment sensor kit for continuous indoor air quality tracking and deployed it in conjunction with established mobile sensing and experience sampling techniques in a cohort study of up to 1584 student participants per data type for 3 weeks at a major research university in the United States. In this paper, we begin by proposing a conceptual framework that systematically organizes human-centric data modalities by their temporal coverage and spatial freedom. Then we report our study design and procedure, technologies and methods deployed, descriptive statistics of the collected data, and results from our extensive exploratory analyses. Our novel data, conceptual development, and analytical findings provide important guidance for data collection and hypothesis generation in future human-centric sensing studies.

\end{abstract}

\section{Introduction}

Human health and behavioral research is primarily conducted in laboratories under conditions that poorly approximate real-world conditions. While this model has been successful, it may miss key aspects of human behaviors that are elicited only during more natural conditions or interactions. This concern has driven interest in developing and using remote sensing technologies to measure individuals completing their normal day-to-day activities in their natural environment. An explosion in modern technologies, many of which are in common usage, now provide the ability to monitor and understand health and human behavior in ways not previously possible. Smartphones, smart home devices, wearables, and online digital behaviors provide new ways to track sleep, emotions, spatial mobility, activity, environmental exposures, and social interactions to name just a few. These technological advances offer opportunities to unobtrusively collect real-time data on a wide range of social-behavioral and health variables with less participant burden and more ecological validity than ever before \cite{harari2016using} 

In the past decade we have seen growing effort worldwide in collecting real-time sensing and experience sampling data from human participants in natural, uncontrolled settings \cite{aharony2011social, wang2014studentlife, stopczynski2014measuring, kiukkonen2010towards, purta2016experiences, boukhechba2018demonicsalmon}. Smartphones are the staple of sensing hardware to measure different aspects of daily behavior \cite{harari2017smartphone}. Four main categories of behavioral patterns can be captured passively: (a) mobility trajectories, measured by GPS and further processed into location clusters that represent significant places visited \cite{canzian2015trajectories}; (b) physical activity, measured by accelerometer and further processed into activity status labels such as walking and staying still \cite{wannenburg2016physical}; (c) social context, reflected in different modes including phone calls, message exchange, and physical proximity detected by Bluetooth, which may be used to reconstruct social networks \cite{wu2018improving}, and; (d) interaction with the device, such as screen unlock status and app usage, which are logged by the smartphone itself \cite{do2011smartphone}. Additionally, ecological momentary assessment (EMA) surveys can be deployed to actively collect participant's self-reports of mood, behavior, and well-being in real time. 

While many aspects of behavior and health can be captured in smartphone sensing data including EMAs with satisfactory ecological validity, we realize that other key dimensions of well-being are better measured using complementary technologies. A person spends a significant proportion of their time at home but measurements of their home environment are not generally investigated in existing studies in parallel with other aspects of daily behavior. To this end, we developed a home environment sensing device, named \textit{BEVO Beacon}, that is capable of continuously collecting and uploading multiple measures of indoor air quality. This device can provide critical insights into a participant's living environment and evaluate its behavioral and health implications. Additionally, sleep is a critical health outcome and independent variable that is difficult to measure objectively using unobtrusive instruments. We argue it is especially beneficial to utilize the sleep measuring capability of wearable devices such as Fitbit to validate EMA answers. 

Furthermore, the vast majority of existing human sensing studies used less than a few hundred participants \cite{cornet2018systematic}. A larger sample is needed to obtain more reliable assessments of the correlations between key behavior and health measures, especially when a large number of variables is assessed. Reflecting these considerations, we conducted a multi-modal human sensing study named \textit{UT1000}, for which we recruited more than 1000 college students as participants over two deployments and distributed a variety of sensors and instruments including smartphone, Fitbit, BEVO Beacon, and EMA. The resulting data allow us to pursue research questions that previous data were unable to accommodate.

With numerous types of technologies and methods potentially available to measure individual humans' health, behavior, and environment, we begin in Section \ref{sec:framework} by proposing a novel conceptual framework that organizes the various modalities of human-centric data based on their properties. Then, we present the study design and the types of data collected in our UT1000 Project in Section \ref{sec:ut1000}. We discuss the procedures and results of our extensive exploratory analyses and visualizations in Section \ref{sec:results}, covering data robustness and novel findings from four types of human-centric data, and caveats from data collection and analysis in Section \ref{sec:discussion}. Through this study we are able to gain comprehensive understandings of the lives of college students and learn valuable lessons about the design, deployment, and data analysis for large-scale human sensing studies. 

\section{Conceptual Framework}\label{sec:framework} 

\begin{figure}
    \centering
    \includegraphics[width = 0.8\columnwidth]{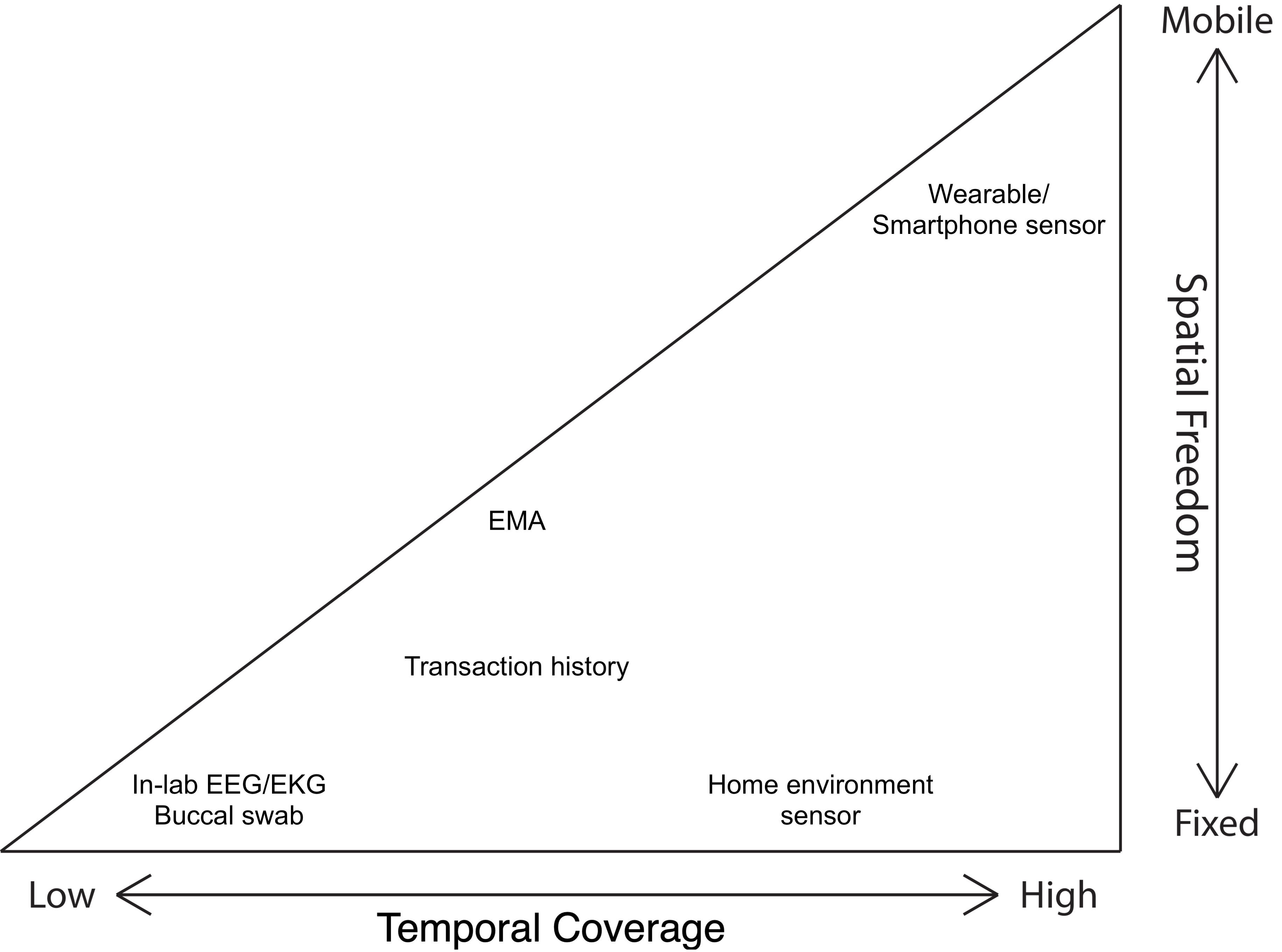}
    \caption{Human-centric Data Modality Framework}
    \label{fig:framework}
\end{figure}

Figure \ref{fig:framework} illustrates the conceptual framework we devised for organizing different technologies and methods for observing human outcomes based on properties of their data collection procedure and resulting data.

A primary property of human-centric data is its \textit{temporal coverage}, represented by the bottom horizontal axis from low (left) to high (right). Temporal coverage is defined by the inherent suitability of a data modality to monitor extended proportions of time of an individual's daily life. Data modalities that can provide only single-time observations are on the low end of the temporal coverage dimension. Examples include (1) traditional survey inventories that are designed to provide a one-time diagnosis of a potential patient and (2) medical procedures that typically require in-person clinical visits such as electroencephalography (EEG), electrocardiography (EKG), and buccal swabs. Some data collection methods or technologies accommodate measurements taken at multiple points in time, thus are placed mid-range along the temporal coverage dimension. Examples include (1) self-reports in response to EMAs delivered via mobile devices and (2) record data such as transaction history which contains user information logged at different times of user engagement with a service or services. Highest temporal coverage is achieved by continuous tracking. Various sensors embedded in devices people carry where they go and install where they stay belong in this category, such as smartphones, wearable devices, and environmental sensors used as smart home technology. Wearable devices including smartphones tend to enjoy an even higher level of temporal coverage than environmental sensors. This is because wearable devices are able to accompany and monitor the user at all times as long as the user keeps the devices on their person and powered on, thus observing the user for longer periods of time compared to environmental sensors, which can only offer \textit{human-centric} sensing capability when the individual in question remains within their proximity (e.g., home environment sensors are only human-centric when the person is at home). 

A second property of human-centric data is its \textit{spatial freedom}, represented by the vertical axis ranging from fixed (bottom) to mobile (top). Spatial freedom is defined by the ability of a data modality to reflect an individual's health, behavior, and environment at a variety of locations. Data modalities with greater temporal coverage tend to permit a wider range of locations where an individual can be monitored, a.k.a. higher spatial freedom. For this reason the coordinate space presented in Figure \ref{fig:framework} is a triangle. A discrete-time measurement of low temporal coverage is typically taken at a specific location, only allowing minimal spatial freedom, thus occupying the left vertex of the triangle. However, spatial freedom may vary greatly among instruments that can track users continuously over time. For example, both a PM2.5 sensor installed at home and a smart wristband can be considered as having high temporal coverage; however, they correspond to fixed and mobile, respectively, on the spatial freedom dimension.

The temporal coverage and spatial freedom of a data modality is often governed by the unobtrusiveness \cite{webb1966unobtrusive} of the technology or method producing it. Data modalities of higher temporal coverage and spatial freedom are usually produced by devices and procedures that are more user-friendly, more portable, and overall less burdensome for the user. Unobtrusive methods allow for more naturalistic and non-interfering ways to monitor a participant's daily life, thus producing measures of greater ecological validity. We identify a major correlation between ecological validity and temporal coverage/spatial freedom: measures of high ecological validity tend to enable observations over extended periods of time and with greater mobility. However, the relation between ecological validity and temporal coverage/spatial freedom is not a necessary one. In the classic example of Barker \& Wright's study of Raymond Birch \cite{barker1951one}, in which a research team followed a subject around for a whole day making observations every few minutes, temporal coverage and spatial freedom are both high however ecological validity is low, because the observation method was extremely intrusive.  

The example data modalities shown in Figure \ref{fig:framework} are based on how the generating methods are naturally and realistically carried out. For example, a buccal swab procedure takes a few minutes to complete but it requires a high level of participation and effort from the patient; therefore even though technically buccal swabs can be frequently administered, we still consider it to be a highly intrusive, one-time measurement. Many human-monitoring technologies, over their course of development, have seen themselves ascend on the temporal coverage scale and often on spatial freedom as well. For example, blood glucose testing used to require clinical visits, thus the burden was high; however, as the technology for on-body continuous glucose monitoring becomes perfected, blood glucose measurements can be obtained with unprecedentedly high unobtrusiveness, allowing its temporal coverage and spatial freedom to increase as well. 

A critical challenge in interdisciplinary research joining social sciences and engineering is the high-fidelity mapping between human-centric constructs and technology-advanced methods. Our framework provides an interface between the constructs and the methods based on their temporal and spatial characteristics and implications for ecological validity and unobtrusiveness. Thus it can help social scientists locate fitting methods to measure the constructs of interest. Furthermore, this framework can be used to guide data collection and hypothesis generation concerning interrelations of different aspects of human behavior. Researchers can delineate a subarea in the triangle to serve as the scope of their own data collection. Hypotheses or research questions can be straightforwardly formulated by linking two spots on the triangle and querying the relationship between the two corresponding data modalities. Moreover, the two-dimensional space describes one individual and can be conceptually stacked up to represent a group of individuals and their data and descriptors. Linkage between slices representing different individuals can inform the generation of research questions into the relations between individual outcomes and inter-personal interactions.  

\section{The UT1000 Project}\label{sec:ut1000}

The UT1000 Project is a multi-modal data collection study conducted at the University of Texas at Austin to measure aspects of the health, behavior, and home environment of a large-scale participant cohort using a wide variety of technologies and methods, including traditional surveys, swabbing, EMAs, smartphone sensing, wearable trackers, and environmental sensors. We undertook two deployments, one in the Fall of 2018 the other in the Spring of 2019, totaling 1584 participants (62\% female) and lasting three weeks each. Participants for this study were recruited through an introductory psychology course. Enrolled students were instructed to sign up for the EMA and smartphone sensing components of the study as a class assignment that counted toward their final grade. Students who did not want to self-track using a smartphone were given the option to record their behaviors and moods by answering emailed EMA questions or keeping a daily diary. The other data modalities such as those from the wearable trackers and environmental sensors, on the other hand, were collected in return for experimental credits which the students used as partial fulfillment of the course requirements. The following subsections outline the different components of the UT1000 Project including the purpose, procedure, and types of data collected. Discussions of different study components are organized into three main categories based on the data modalities, namely \textit{single-time}, \textit{multiple-time}, and \textit{continuous} measures, following an order of temporal coverage from low to high consistent with the horizontal dimension of Figure \ref{fig:framework}.

\subsection{Single-time measures} 
\subsubsection{Home Environment and Health (HEH) Survey}\label{subsubsec:heh}

The Home Environment and Health (HEH) questionnaire consists of 63 questions asking students to report on home environment factors such as their current living situation including number of roommates, number and type of pets, and flooring type; their recent health and medical histories including colds, allergies, and flu shots received; and other behaviors such as hand washing frequency and use of electric scooters. A full list of the HEH questions are provided in Appendix A. The purpose of this survey was to obtain a better understanding of the participants' home environment and to clarify discrepancies found in the other data streams.

The HEH questionnaire was a voluntary survey sent directly to participants via the email they provided to register for the study. Completing the HEH survey was a prerequisite for the subsequent home environment sensing component of the study discussed in Section \ref{subsubsec:beacon}. The survey was sent once during the first two weeks of the study period. Participants were asked to fill out the survey based on their situation when they received it rather than some time in the past or the future. A total of 56 participants completed the HEH questionnaire, with 46 in Fall 2018 and 10 in Spring 2019. 

\subsubsection{Student Environment and Buccal Swabbing}\label{subsubsec:swab}

A subset of the study participants were provided with a dust sampling kit to collect dust samples from various surfaces in their home and classroom environment. The same participants that completed the HEH survey were given the kits ($N=56$). The kit consisted of six, individually-packaged Phosphate-Buffered Saline Tween-20 (PBST) wetted FLOQswabs\textsuperscript{\textregistered} (manufactured by COPAN\footnote{\url{https://www.copanusa.com/sample-collection-transport-processing/floqswabs/}}, Murrieta, CA) and six corresponding plastic, resealable test tubes that participants would place the swabs in after collecting samples. Participants were asked for identification in order to gather sampling materials from a refrigerator-equipped storefront created ad hoc in a convenient place at a central location of the university. Testing materials to be distributed to different participants were labeled with distinct barcodes so that we could easily trace the materials back to the participants and streamline the checkout process.

Participants followed instructions to collect samples from the interior and exterior of their front door trim, cellphone screen, living room floor, HVAC filter or air diffuser if applicable, and a desktop where they normally sit when attending university classes. After sample collection, participants sealed swabs in the provided test tubes and placed them in their refrigerator until transportation to the university lab. When participants return the testing materials, material barcodes were scanned, the identity of the participant cross-referenced to the materials, and the temperature-sensitive samples were stored until transfer to a -4$^\circ$F freezer daily after storefront closure. They were asked to provide feedback on the challenges while performing home sampling and also whether they are willing to submit a buccal swab. If they consented, the research assistant in charge of operating the storefront would ask the participant to use a swab to collect a sample from the inside of their cheek. Samples were then stored in a small, resealable test tube and refrigerated before transfer to a -80$^\circ$F daily after storefront closure. 

The dust samples are useful to help understand more deeply about the participants' home environment beyond the HEH survey and what they might be exposed to on campus when attending classes. Examination of the dust samples can determine what types of microbial exposures commonly occur in students' indoor environments. Buccal swabs can be used for a variety of reasons, but were conducted as part of this study to understand how certain chemical markers like cytokine levels are related to mood and stress in participants.

\subsection{Multiple-time measures}
\subsubsection{Ecological Momentary Assessment}\label{subsubsec:ema}

Ecological Momentary Assessments (EMAs) involve brief questions about a participant's behavior and feelings that are answered in real-time while the participant is in their natural environment. EMAs were administered using the Beiwe mobile application\footnote{\url{https://www.hsph.harvard.edu/onnela-lab/beiwe-research-platform/}} running on their smartphones at regularly scheduled times throughout each day. For both the Fall 2018 and the Spring 2019 cohorts, EMAs were drawn from four categories of questions: sleep questions, momentary context questions, momentary well-being questions, and an audio question. The full text of these questions are included in Appendix B. Briefly, the three sleep questions were designed to assess the duration and quality of sleep, momentary context questions were designed to determine what the participant was doing and who they were doing it with, and five well-being questions sampled participants' mood (sadness, loneliness, contentment, and stress) and energy level on a Likert scale. The audio question asked the participant to describe what they were doing and to include a brief segment of background noise. 

Both the Fall 2018 and Spring 2019 cohorts received EMAs at five different times during the day. At 9am each morning they received the three sleep questions, four momentary context questions that were framed to assess the 15 minutes prior to receiving the EMAs, and five mood questions that were framed to assess how they feel at the moment. At 12pm, 3pm, and 6pm participants received EMAs that were similar to those from the morning, except that the sleep questions were removed and the audio question was added. Each night at 9pm they received the four momentary context questions and five mood questions that were framed to assess participant behaviors and feelings across the entire day.

\subsection{Continuous measures} 
\subsubsection{Smartphone}

Passive monitoring data were collected using the Beiwe digital phenotyping platform, which is a freely available open-source system that includes mobile phone applications for Apple iOS and Google Android operating systems, and a backend server implemented in Python. The backend server was run using Amazon Web Services cloud based computing infrastructure. The backend includes a study administration web application for designing and conducting Beiwe studies and monitoring their progress, an API for sending study parameters to and receiving data from the mobile phone applications, a database for storing study state information, and an encrypted Amazon Simple Storage Service bucket for storing study data. The Beiwe developers maintain versions of the app in the Apple App store and Google Play store for easy deployment to study participants.

The full set of passive monitoring parameters are listed in the Appendix C. Due to differences in operating system security settings and device capabilities, different data sources were collected on Android and iOS devices. Basic device and operating system information (make, model, version), accelerometer, GPS, and power state data were available and collected on both devices. iOS specific data sources include gyroscope, magnetometer, the proximity of the device to the user, and whether the phone is connected to the internet by WiFi or cellular. Android specific data sources include a list of WiFi routers and Bluetooth devices in the phone's proximity, the time, duration, and hashed phone numbers for incoming and outgoing calls, and time, message length, and hashed phone numbers of incoming and outgoing text messages. To maintain participant privacy, WiFi and Bluetooth identifiers, and phone numbers are encoded with a hashing function. The function is unique however, so calls to the same destination and proximity to the same WiFi access points can be tracked across time. 

The Beiwe mobile application was configured to store collected information locally and to upload it only when connected to the internet using WiFi. If an error was encountered during transmission, the app stores the data and retries transmission until receiving an indication that the data was successfully received by the backend. Each data source is stored in its own set of CSV files that are broken down and organized by timestamp. These files are encrypted on the phone before being transmitted over an SSL connection to the backend. When received, the data are unencrypted, processed to correct for errors, and update received data statistics, and then re-encrypted for storage. All encryption is performed using randomly generated participant specific encryption keys. 

As per the study design, participants were instructed to download and allow all permissions for the Beiwe platform. Each participant had a randomly generated identification tag that consisted of eight letters and numbers and were prompted to create their own password after entering a temporary password given to them by the study coordinator. Participants did not have direct access to their data and used the login credentials when completing the EMA surveys. 

\subsubsection{Wearable Activity Tracker} 

Participants' activity and sleep patterns were captured using the Fitbit Charge2\textsuperscript{TM} wearable activity trackers. The devices require participants to input their height, weight, gender, and age to accurately calculate the number of steps taken, calories burned, and the wearer's heart rate. In addition, participants can track different exercises by selecting them from the device's interface or through the paired smartphone application. Most Fitbit products, including those supplied to the study participants, are capable of passively monitoring the wearer's sleep as long as the device detects that the user has been asleep for a minimum number of hours. The wearer's sleep is subdivided into four categories based on movement and heart rate: awake, light, deep, and rapid-eye-movement (REM). Over the past few years, many studies have looked at the accuracy and utility of using Fitbit and other personal monitoring devices in sleep studies \cite{liang2018validity,de2018validation,baron2018feeling,liang2019accuracy}. Results from these studies show that these devices can be useful when determining total sleep time, awake time, and the amount of time spent in REM sleep.

Similar to the swabs discussed in Section \ref{subsubsec:swab}, each Fitbit was numbered, barcoded, and provided to participants also at the storefront location if they consented to participate in the activity monitoring component of the study. In addition to receiving the device, each participant was required to register a Fitbit account and download the smartphone application if they did not already hold an account. Participants were asked to wear the activity monitors as much as possible, removing them only when bathing, participating in aquatic activities, or charging the device. Participants were instructed to wear the Fitbit monitor over a period of at least two weeks and were free to use the device outside of the study requirements. If broken or damaged, participants were given a new device to register and use for the remaining portion of the study.

\subsubsection{Building Environment and Occupancy Beacon}\label{subsubsec:beacon} 

\begin{figure}[]
    \centering
    \includegraphics[width=0.5\columnwidth]{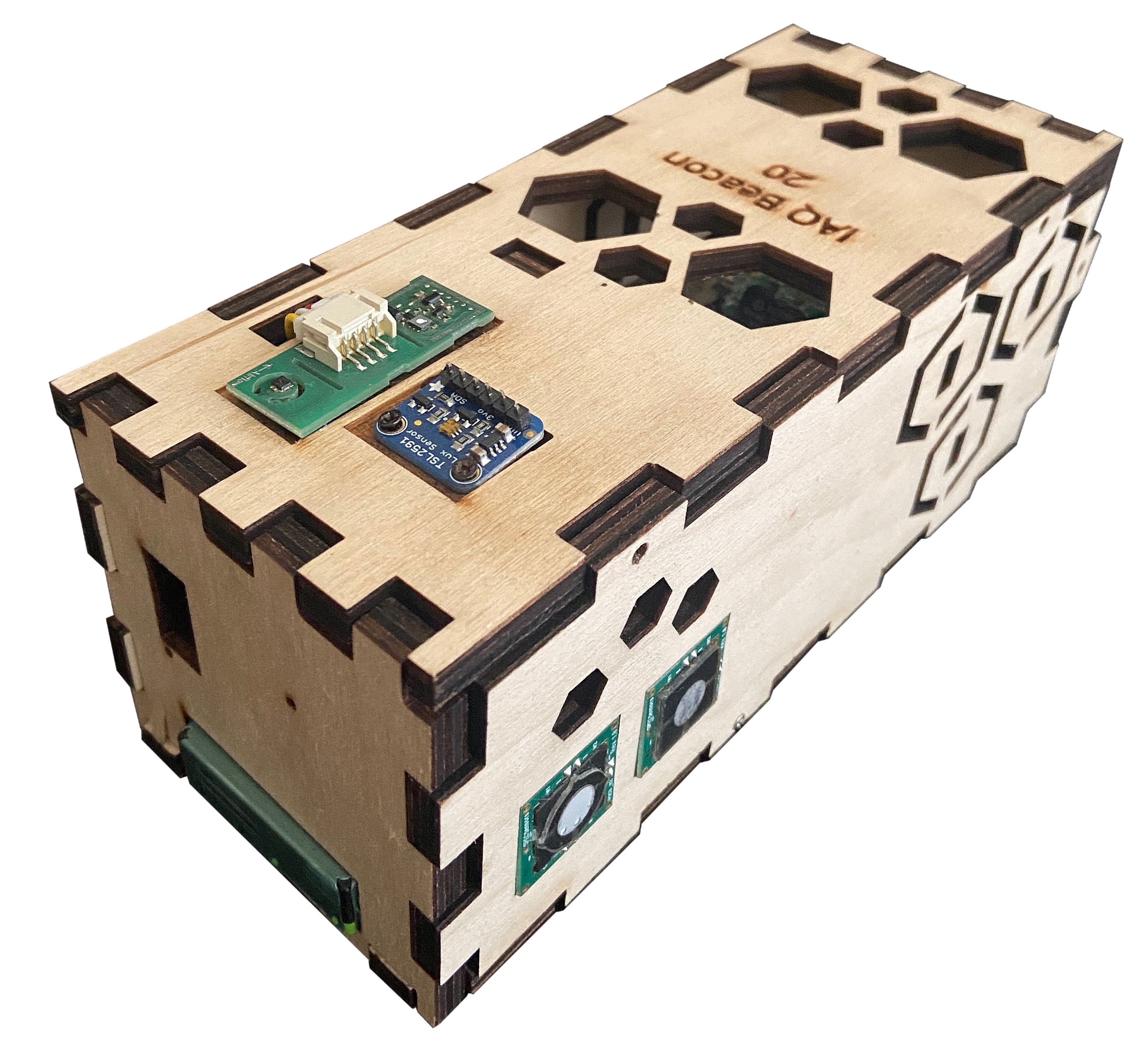}
    \caption{Building EnVironment and Occupancy (BEVO) Beacon}
    \label{fig:beacon_portrait}
\end{figure}

The Building EnVironment and Occupancy (BEVO) beacon is a low-cost sensor platform we developed in-house that is capable of measuring multiple indoor environmental quality (IEQ) variables in addition to detecting Bluetooth and WiFi signals. The BEVO Beacon consists of a Raspberry Pi (RPi) micro-computer connected to a variety of environmental sensors arranged into a 6"$\times$6"$\times$4" housing made of plywood and acrylic (see Figure \ref{fig:beacon_portrait}). The RPi can detect Bluetooth devices and WiFi access points in its proximity, which can help determine occupancy. The RPi is additionally capable of storing data captured by itself and the co-locating IEQ sensors on a local micro SD card, and then uploading the data to a cloud-based storage system hosted by the Texas Advanced Computing Center once connected to WiFi. Table \ref{tab:bevo_variables} outlines the environmental sensors used in BEVO Beacon and the variables they measure. The RPi is housed in a lower compartment, separated from the sensors and wired to a small fan to help with heat regulation. To avoid the possibility of re-sampling air trapped inside the device's housing, and further help with heat management, the environmental sensors are housed in a compartment above the RPi with their inlets/exhausts exposed to the ambient air. The BEVO Beacon requires a 5B micro-USB portal connection to power.  

\begin{table}[]
    \centering
    \begin{tabular}{p{0.23\linewidth}p{0.14\linewidth}p{0.12\linewidth}p{0.39\linewidth}}
        \toprule
        Variable & Unit & Sensor & Notes\\
        \midrule
        Temperature (T) & $^{\circ}F$ & Adafruit SHT31-D & Occupant thermal comfort \\         \midrule
        Relative humidity (RH) & \% & Adafruit SHT31-D & Occupant thermal comfort \\ \midrule
        Particulate matter (PM) & $\mu g/m^3$ & Plantower PMS5003 & EPA-specified criteria air pollutant with respiratory health implications \\    \midrule
        Total volatile organic compounds (TVOCs) & $ppb$ (parts per billion) & Adafruit SGP30 & Compounds with a wide-range of health implications from respiratory issues to cancer\\ \bottomrule
    \end{tabular}
    \caption{IEQ variables measured by the environmental sensors housed in the BEVO Beacon}
    \label{tab:bevo_variables}
\end{table}

The commercially available sensors used on the BEVO Beacon afforded us two benefits. These sensors were low-cost ($<$US\$100), which allowed us to develop and deploy more devices than what is typically done when measuring indoor air quality. The second benefit is that these sensors serve as the base units in many other commercially available IEQ products. Using the base units rather than off-the-shelf devices ensures that there is no proprietary algorithm that alters the raw values measured by the sensors. However, reliability and accuracy are two main issues surrounding the use of low-cost sensors since they use less sophisticated electronics than high-grade reference monitors that can cost more than US\$10000. Our beacon development effort is a step toward integrating low-cost sensors to achieve high home-sensing performance. 

BEVO Beacons were each assigned a number and a barcode. Only the participants who completed the HEH survey (see Section \ref{subsubsec:heh}) were eligible to receive a BEVO Beacon. Participants who were eligible and consented to participation were instructed to stop by the storefront and check out the device. At check-out, the device was supplied to the participant along with a 5V micro-USB to wall outlet adapter. Participants were instructed to power the device using any open outlet in their home. In total, we distributed 15 BEVO Beacons: five in Fall 2018 and ten in Spring 2019.

\section{Exploratory Results}\label{sec:results}
\subsection{Data collected}\label{subsec:data_collected}

\begin{figure}[]
    \centering
    \includegraphics[width=0.8\columnwidth]{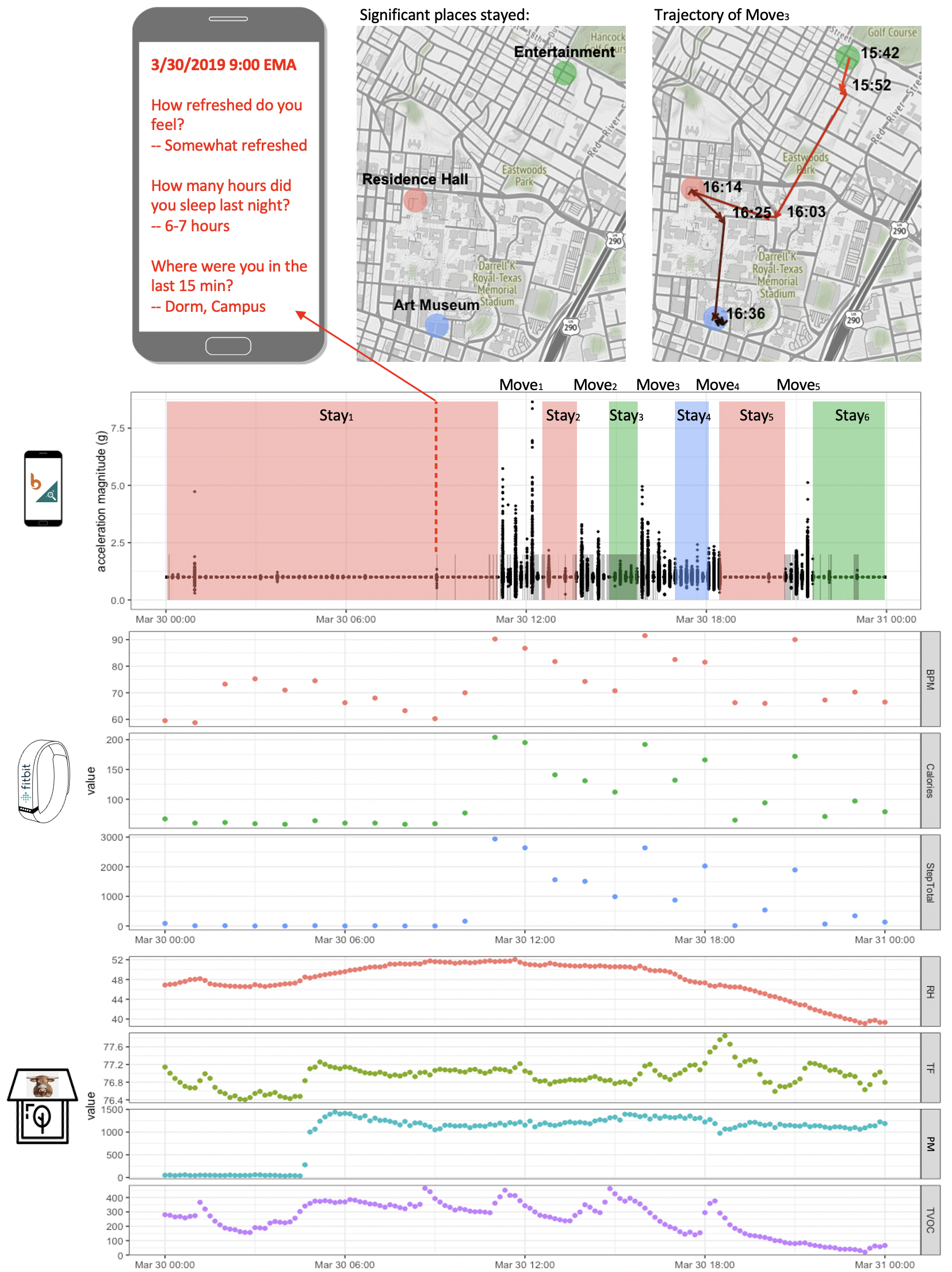}
    \caption{Data collected from the smartphone, Fitbit, and BEVO Beacon of an example participant during a given day (March 30th 2019). Plotted data modalities are: EMA (questions and answers shown against phone image background on top), GPS (clustered significant places and an example movement trajectory shown in maps on top and as vertical bands), accelerometry (black dots), screen activity (short grey bands), heart rate (BPM), calories spent in the past hour (Calories), steps taken in the past hour (StepTotal), home relative humidity (RH), home temperature in degrees Fahrenheit (TF), home particulate matter in $\mu g/m^3$ (PM), and home TVOC (TVOC).}
    \label{fig:example_pid}
\end{figure}

To showcase the diverse data streams collected from our participants, we plot in Figure \ref{fig:example_pid} data collected from a particular participant's smartphone, Fitbit, and BEVO Beacon on a given day (March 30th, 2019) as an example. The top third of Figure \ref{fig:example_pid} shows the data types collected by the Beiwe smartphone platform. We conducted temporal clustering \cite{kang2005extracting} with the raw GPS traces and processed them into periods of stay at significant places (represented by the colored vertical bands) and periods of movement between significant places (represented by the white spaces between the colored bands). We used Open Street Map API to query for the place type of the significant places found. We show the geographic location and venue type of the significant places, as well as the trajectory and time of the third transition period of the day (``Move${}_3$") in the two maps on top. In total on March 30th, 2019, the participant made 6 stays at 3 distinct places (residence hall, art museum, and entertainment venue) and made 5 trajectories of movement between them. The participant spent her entire night and most of the morning (midnight to around 11am) at the residence hall (red band/dot), which was her main residence. Notice that even though the trajectory of Move${}_3$ (top right map) passes through the residence hall but did not register another period of stay, suggesting that the participant merely swung by the residence hall without making a stop for an extended period of time. 

\begin{figure}[]
    \centering
    \includegraphics[width=\columnwidth]{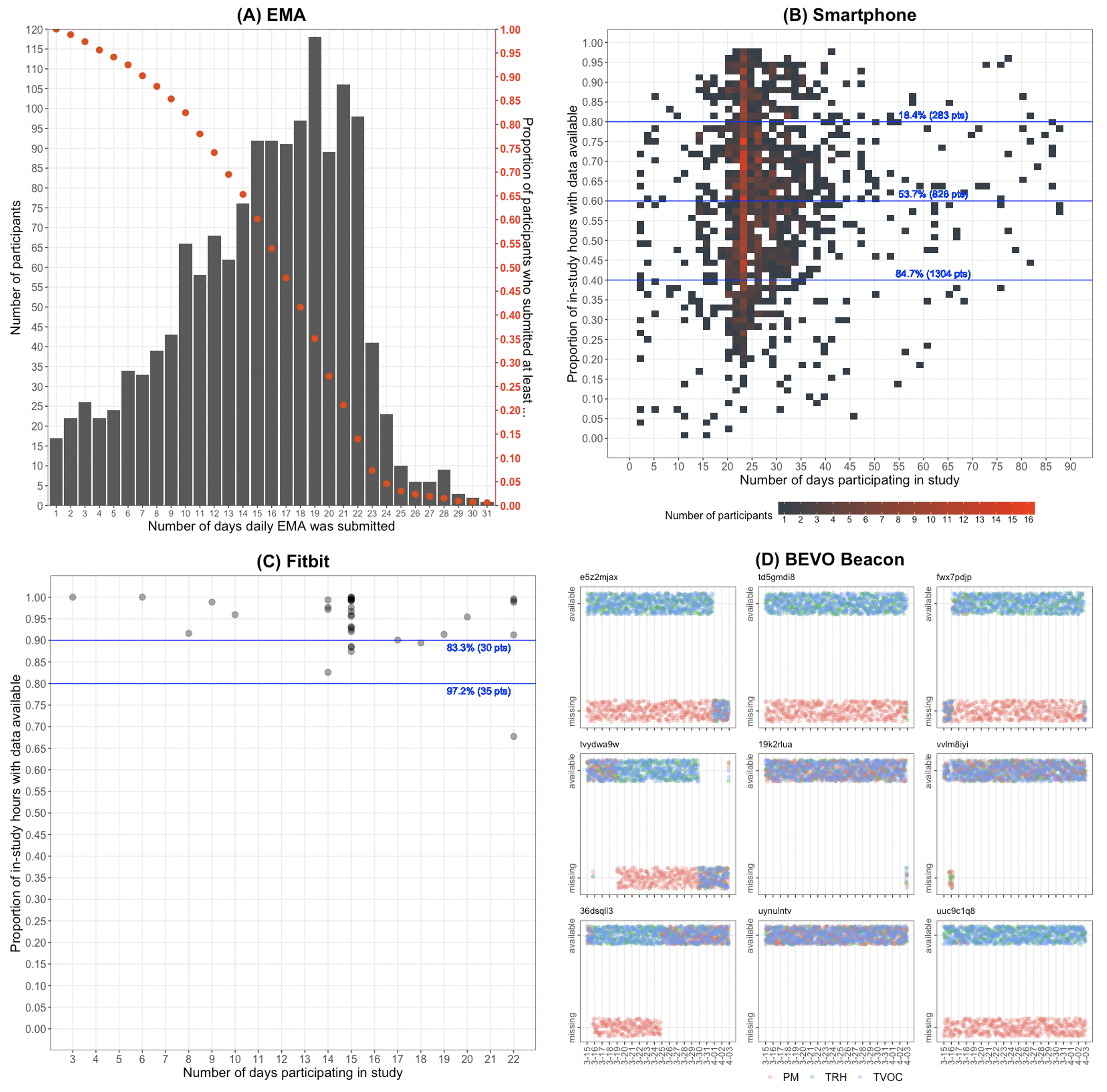}
    \caption{Completeness of four types of data collected from participants in the Fall and Spring deployments combined: (A) daily EMA (1482 participants); (B) Smartphone sensing, GPS data shown as example (1539 participants); (C) Fitbit (36 participants), and; (D) BEVO Beacon (9 participants).}
    \label{fig:data_reliability}
\end{figure}

During the three-week official study period, we were able to record some GPS and accelerometer data from approximately 950 participants each day. Within these days, data completeness for GPS and accelerometer followed a highly regular and well-synced daily cycle where the least percentage of participants submitted data in the early morning hours and the highest during the evening hours. Averaging the hourly completeness percentages gives us a daily average of around 65\% of the participants submitting data during any given hour. In addition to mobility information, smartphone acceleration magnitude is plotted as black dots and episodes of unlocked screen as short grey bands. We observe that periods of high acceleration magnitude correspond well with periods of movement between places. Screen activity, on the other hand, varies heavily depending on the place: for example, the screen stayed unlocked during the entire Stay${}_3$ at the entertainment venue, but locked during Stay${}_4$ at the art museum. Moreover, the participant responded to an EMA survey at 9am, providing a self-report of her sleep quality, hours of sleep, as well as the semantics of her location at the moment. The participant did not respond to any EMA questions scheduled at other times of the day. Her answer ``dorm; campus" to the location question matches with the significant place detected by GPS. 

The middle third of Figure \ref{fig:example_pid} shows three data streams recorded by Fitbit: heart rate (BPM), calories expense (Calories), and steps taken (StepTotal). The patterns of fluctuation of the three data streams are largely in sync with one another, with values significantly higher during the day than during the night. Three points of peak values in heart rate and calories (11am, 4pm, 9pm) visibly correspond to time intervals of high smartphone accelerometer readings, suggesting a positive correlation between smartphone accelerometry and physical activity status recorded by wearable devices. The bottom third of Figure \ref{fig:example_pid} shows four data streams recorded by BEVO Beacon, all of which are metrics of indoor air quality. Note that only when the participant is located at her home location (indicated by red band in this plot, specifically Stay${}_{1,2,5}$) is she exposed an environment described by these metrics; when she goes away, the metrics merely reflect her home environment status that does not affect her directly. We observe a sharp rise in humidity and PM concentration between 4-5am, which is potentially indicative of a change in the HVAC system. 

\subsection{Data completeness}\label{subsec:data_completeness}

Completeness encompasses two metrics: first, the amount of time each participant stays in the study and continues to submit data, actively and passively; second, when the data type is continuous, the proportion of time during the entire period of participation that data are available. The first metric is important because it represents \textit{participant compliance} and a higher value in it (or, closer to the length of the intended study period) indicates more successful participation. The latter measure is also important in that it represents \textit{data continuity} and a higher value in it indicates fewer data-missing intervals during the total period of time a participant is submitting data. 

In Figure \ref{fig:data_reliability} we present the completeness of four major types of data we collected, namely EMA, smartphone, Fitbit, and BEVO Beacon. In Figure \ref{fig:data_reliability}-(A), we show the number of participants (height of bars) who submitted answers to daily EMAs for all different numbers of days (horizontal position of bars), as well as the percentage of participants (projection of red dots on vertical axis on right) who submitted answers to daily EMAs \textit{at least} a certain number of days (horizontal position of red dots). We observe that more than 60\% of participants submitted daily EMA answers for more than 14 days and more than 20\% more than 21 days. 

In Figure \ref{fig:data_reliability}-(B), we show the distribution of participants with respect to the total duration of time they were contributing smartphone GPS data (i.e., compliance, on the horizontal axis) and the proportion of that duration for which their data are available (i.e., continuity, on the vertical axis). We use the color of square cells to indicate the number of participants that fall in particular compliance-continuity boundaries: the brighter-colored vertical bar between 20-25 days correspond well with our planned study length which is three weeks. Shown by the blue horizontal lines, 283 participants (18.4\%) had smartphone GPS data available for more than 80\% of the hours they participated in study, 53.7\% of participants more than 60\%, and 84.7\% of participants more than 40\%. 

Figure \ref{fig:data_reliability}-(C) contains the same information as (B) but plotted for Fitbit data. As opposed to aggregating the number of participants like we did for (B), because of the lower number of participants who chose to wear a Fitbit (36), we simply represented each Fitbit participant as an opaque grey dot, forming a darker cluster when more participants fall in a close region of compliance-continuity value combination. Compared to smartphone data, Fitbit data enjoyed significantly higher continuity: out of the 36 participants, 30 submitted data for more than 80\% of their in-study time. This may suggest that Fitbit as a wearable device requires less human attention and interference, as opposed to smartphones which are constantly being handled and require more frequent charging, and thus are more prone to produce uninterrupted data streams. 

Out of the 15 BEVO Beacons we distributed, we found that only nine recorded and uploaded data reliably and they were all from the Spring 2019 deployment. Due to such a small number of participants who returned substantial environmental sensing data, in Figure \ref{fig:data_reliability}-(D) we simply plotted entire time series of data availability of each IEQ measure captured by BEVO Beacons (indicated by different colored dots, jittered) for each of the nine participants. Visibly, three of the nine BEVO Beacons submitted data perfectly whereas the remaining six primarily had trouble submitting PM data (red strip on the bottom). Upon inspection we found that communication errors between the sensors and the RPi accounted for a large amount of data loss. Fluctuations in power delivery between the RPi and the environmental sensors caused some environmental sensors to go offline for periods and resulted in data loss during the study. 

One highly notable difference is in the number of participants who signed up for smartphone sensing compared to wearable and home environment sensing ($>$1000 vs. $<$100). We believe that both hardware availability and the incentive structure contributed to this discrepancy. First, the smartphone component of the study required the participants' own primary phone, which was widely available; whereas the Fitbit and BEVO Beacon components required extra hardware we needed to purchase, build, and provide for the participants, thus limiting the number of participants we could enroll. Second, the incentive for students to complete smartphone sensing and EMAs was to receive credits on an assignment that counted toward their final grade (with the alternative to opt out of using smartphone and only logging manually instead), which proved to be an effective strategy to improve participant compliance \cite{harari2017evaluation}; whereas participation in the other parts of the study was rewarded by extra experimental credits, which may not have been nearly as attractive to the students. 

\subsection{EMA findings}\label{subsec:ema_findings}

\begin{figure}[]
    \centering
    \includegraphics[width=\columnwidth]{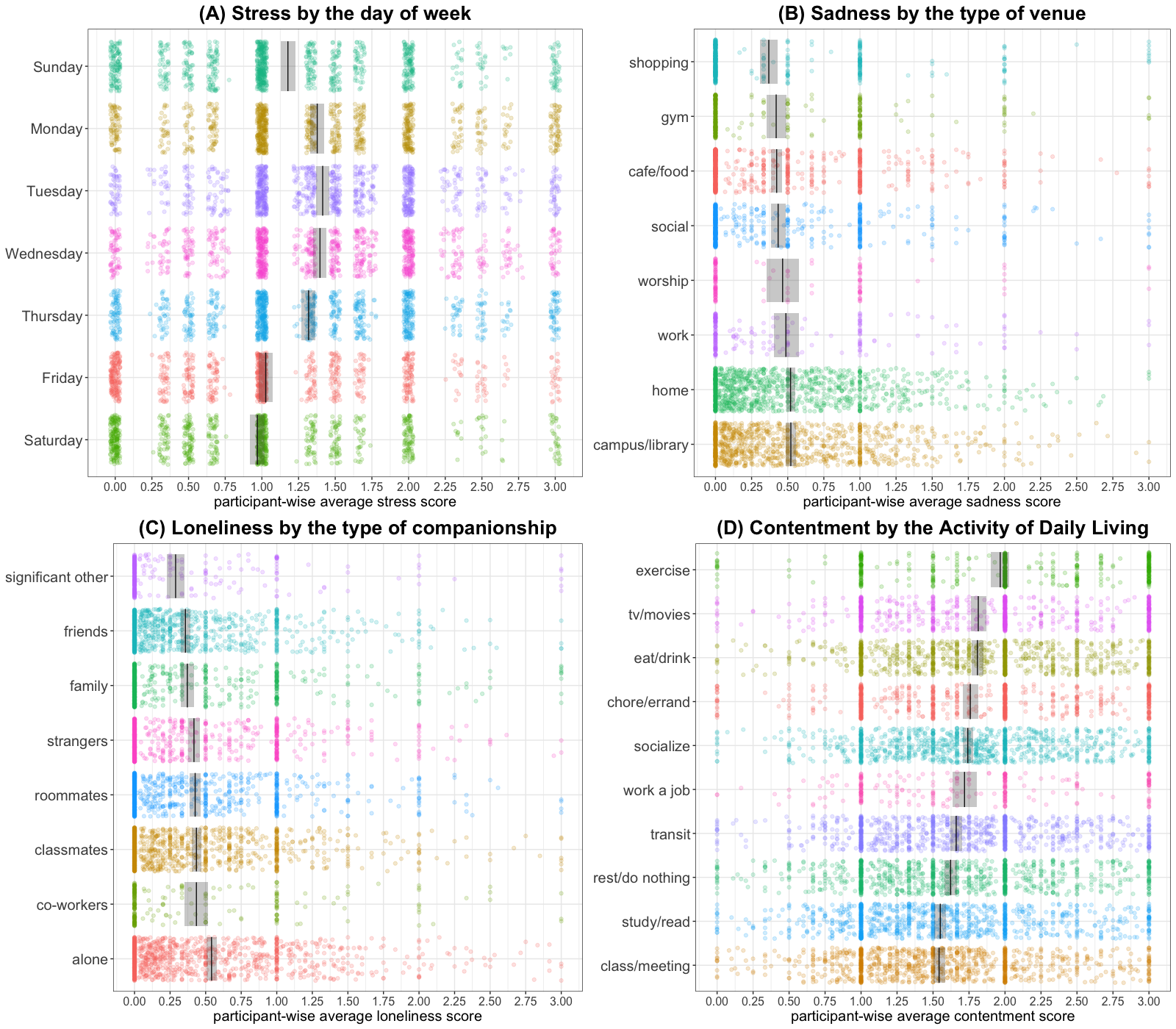}
    \caption{Self-reported momentary mood ratings from Ecological Momentary Assessments. The type of venue, type of companionship, and the Activity of Daily Living were self-reported together with the mood surveys. Each dot represents the average value of a participant under a corresponding category. The black vertical bars indicate the mean of the participant-wise (i.e., calculated within each participant) average scores and the surrounding grey bands indicate a bootstrapped 95\% confidence interval of the mean. In panels (B)(C)(D), categories are arranged from bottom up in a least-to-most-desirable order based on the corresponding mood outcome.}
    \label{fig:ema_findings}
\end{figure}

With our EMA data we explore how participants' self-reported momentary mood ratings vary under different temporal, spatial, social, and behavioral contexts. Each context is specified by a categorical variable with multiple levels. For temporal context, we focused on the day of week and thus have 7 levels. For spatial context, we organized the answers to the place type EMA question into 8 categories: \textit{campus/library}, \textit{home} (including dorm), \textit{work} (the site of a job), \textit{worship} (e.g., church), \textit{social} (Greek house, friends' house, or party place), \textit{cafe/food}, \textit{gym}, \textit{shopping} (store or mall). For social context, we organized the answers to the ``who are you with" EMA question into 8 categories: \textit{alone}, \textit{strangers}, \textit{classmates}, \textit{family}, \textit{co-workers}, \textit{roommates}, \textit{friends}, \textit{significant other}. For behavioral context, we organized the Activity of Daily Living question into 10 categories: \textit{class/meeting}, \textit{study/reading}, \textit{rest/do nothing}, \textit{transit}, \textit{work a job}, \textit{socialize} (including talk, text, and using social media), \textit{chore/errand}, \textit{eat/drink}, (watching) \textit{tv/movies}, and \textit{exercise}. For each of the four mood outcomes (sadness, stress, loneliness, contentment), the self-ratings were evaluated on a 0-3 ordinal scale, with 0-3 corresponding to a mood outcome being \textit{not at all}, \textit{a little bit}, \textit{quite a bit}, and \textit{very much} so respectively. We computed a participant-wise average for each participant when they are under each of the contextual categories listed above. We bootstrapped the mean of the participant-wise averages for each contextual category with 2000 with-replacement samples to determine a 95\% confidence interval. We show some of the results in Figure \ref{fig:ema_findings}. 

Figure \ref{fig:ema_findings}-(A) shows participants' stress level on different days of week. A participant tends to experience significantly lower stress on Fridays and Saturdays than the rest of the week while Monday to Wednesday are the most stressful days. Figure \ref{fig:ema_findings}-(B) shows participants' sadness level felt at different kinds of places. A participant tends to feel the least sad at a store or mall and saddest at home or on campus. Figure \ref{fig:ema_findings}-(C) shows participants' loneliness while accompanied by different types of individuals. A participant tends to feel the most lonely being alone and least lonely being with a significant other; and there appears to be a decrease in loneliness with the type of companionship becoming more intimate. Figure \ref{fig:ema_findings}-(D) shows participants' contentment experienced while engaging in different types of daily activities. School related activities such as attending classes and studying see the lowest contentment whereas exercise is associated with the highest contentment, with watching tv/movies and dining being not-surprising runner-ups. These results portray the general patterns of a college student's psychological experience in her daily life. 

\begin{figure}[]
    \centering
    \includegraphics[width=0.66\columnwidth]{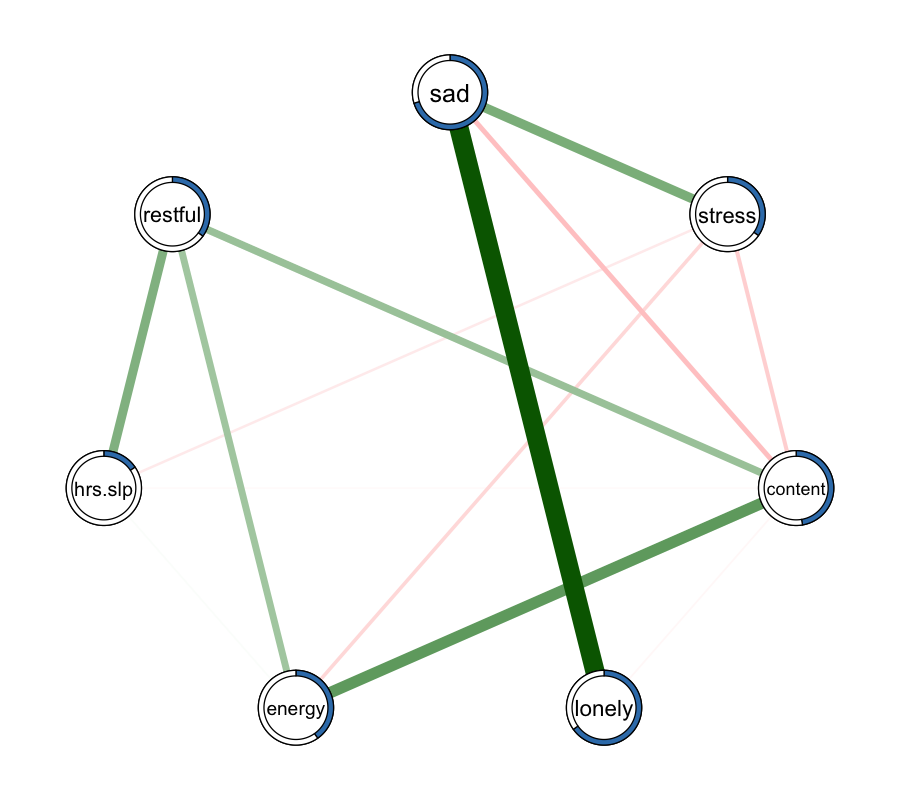}
    \caption{Interdependency between a participant's self-reported average daily mood ratings (sadness, stress, loneliness, contentment), energy level (energy), sleep quality (restful), and sleep duration (hrs.sleep). For each variable (node), a LASSO regression model is fitted with other variables as predictors and the regularization parameter $\lambda$ optimized by 10-fold cross validation. The completeness of the ring around each node indicates the proportion of its variance explained. The color and the weight of the edges are determined by the sign (green-positive; red-negative) and the value of the coefficients from the LASSO models fitted.}
    \label{fig:ema_interdependency}
\end{figure}

In addition to the effect of daily life context on mood outcomes, we further explore the interdependency between these outcomes together with other health outcomes reported by the participants, that is how these outcomes are robustly related with one another. We focus on 7 daily measures, namely sadness, stress, loneliness, contentment, energy level, restfulness of sleep, and hours of sleep. The values of these outcomes are obtained from the daily EMA questions: end-of-day self-assessment of mood and energy level experienced during the day, and beginning-of-day self-assessment of the quality and duration of the previous night's sleep. We start by taking the average of each participant's 7 daily measures (so that each data point represents a distinct participant and is independent) and then fit a LASSO regression model for each of the 7 measures using the other 6 measures as predictors, with the regularization parameter $\lambda$ optimized by 10-fold cross validation (so that unimportant and fortuitous correlations get reduced to zero). Through this operation (essentially a double round robin), the relation between each pair of outcomes receives a coefficient in two models. We take the average of the two coefficients and use that value as the dependency value between the corresponding pair of outcomes. Figure \ref{fig:ema_interdependency} shows these dependency values by the edges between nodes that represent the daily outcomes: the dependency values' sign corresponding to the color (green being positive) and magnitude proportional to the thickness. We also computed the percentage of variance in each outcome explained by the other outcomes via the fitted models and indicate as the completeness of the ring surrounding each node in Figure \ref{fig:ema_interdependency}. 

The most prominent dependency we observe is between sadness and loneliness, with sadness robustly correlated with only loneliness, stress, and contentment in the expected directions, but neither with energy level nor with sleep quality and duration. 70.4\% of variance over participants' average daily sadness levels can be explained by their other three mood outcomes and that is the highest percentage of all 7 outcomes investigated. We observe a strong correlation between perceived sleep quality (restful) and perceived sleep duration (hrs.sleep), but also a clear difference between how the two sleep metrics are correlated with other measures. Perceived sleep quality of the previous night is robustly correlated with the contentment and energy level of the current day whereas sleep duration does not display such effect. Of the four mood outcomes, contentment appears to the only one that has a strong correlation with both energy level and the sleep quality of the previous night.

\subsection{Smartphone findings}\label{subsec:smartphone_findings}

\begin{figure}[]
    \centering
    \includegraphics[width=\columnwidth]{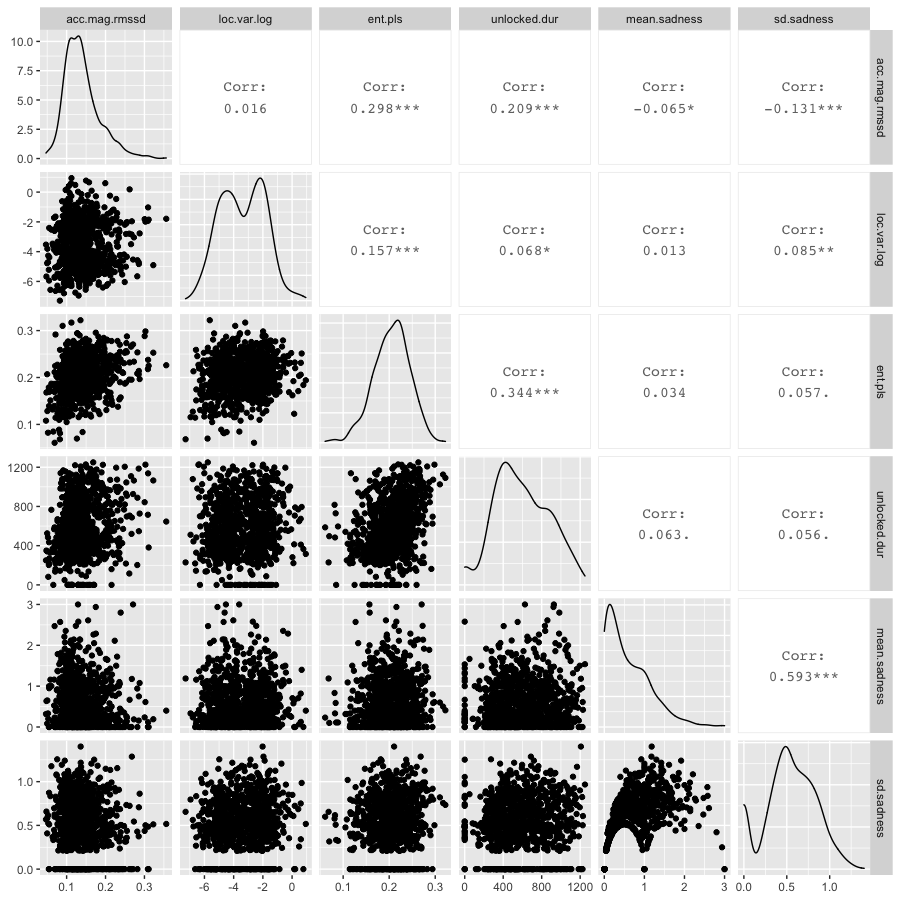}
    \caption{Scatter plot matrix of four mobile sensing phenotypes and two mood symptom metrics. Density plots of each variable are shown on the diagonal. Pairwise correlation values together with their significance level are shown in the cells above the diagonal (p-value: \textbf{***}$<0.001$; \textbf{**}$<0.01$; \textbf{*}$<0.05$*; \textbf{.}$<0.1$)}
    \label{fig:beiwe_findings}
\end{figure}

Smartphone sensing data has been extensively studied for its utility in diagnosing and predicting mental and physical health symptoms. Existing research has identified a number of digital phenotypes \cite{rohani2018correlations} extracted from GPS, accelerometer, phone usage, and phone call/SMS log data that are significantly correlated with aggravated mood. Most of these studies used smartphone data collected from less than 200 participants and we wonder whether some of the important correlations can be replicated in our data with $\sim$1500 participants. We focus on accelerometer, GPS, and screen usage since these three types of smartphone data are available for both iPhone and Android using participants, the latter constituting 1/8 of the entire cohort. We selected from our participants those who completed at least 7 days of daily sadness level self-reports and computed the average daily value of the following four features: \textit{acc.mag.rmssd}, \textit{loc.var.log}, \textit{ent.pls}, and \textit{unlocked.dur}. \textit{acc.mag.rmssd} refers to the root mean square of successive differences of acceleration magnitude (unit: gravity) and quantifies the suddenness of the smartphone's movement with a higher value indicating more transition between staying relatively still to moving intensely. \textit{loc.var.log}, log-transformed location variance, is computed by summing the variance in longitude and latitude values of a participant's GPS coordinates and then taking the natural logarithm. Participants who visit many different places that are far apart register a higher value in location variance. \textit{ent.pls} is defined by the normalized entropy of the amounts of time spent at difference places. To compute this feature, we first undertake a temporal clustering procedure mentioned in Section \ref{subsec:data_collected} to obtain a list of key places, then find the duration of time spent at each of the key places, and finally calculate the normalized entropy over these duration values. Lastly, \textit{unlocked.dur}, a.k.a. duration of screen unlocked, is simply the amount of time (minutes) a participant's smartphone screen stays unlocked. In addition to these 4 smartphone phenotypes, we computed \textit{mean.sadness} and \textit{sd.sadness}, the mean value and standard deviation of daily sadness scores representing the severity and the fluctuation intensity of sadness respectively, to serve as the health outcomes of interest. 

Figure \ref{fig:beiwe_findings} shows pairwise scatter plots and correlations among the four smartphone sensing features and two mood symptom metrics. \textit{mean.sadness} and \textit{sd.sadness} are highly correlated, suggesting that participants who reported greater sadness during the study period also reported greater variation of sadness levels. Apart from the correlation with each other, both \textit{mean.sadness} and \textit{sd.sadness} appear significantly and negatively correlated with \textit{acc.mag.rmssd}, indicating a positive connection between active motion and improved mood. However, they do not display a significant correlation with other sensing features, with the exception of \textit{loc.var.log} which shows a significant positive correlation with \textit{sd.sadness} but none with \textit{mean.sadness}. The three variables, \textit{acc.mag.rmssd}, \textit{unlocked.dur}, and \textit{ent.pls}, are all highly correlated with one another. This suggests that the participants whose smartphones have more highly fluctuating movement tend to divide their time more evenly at different places and spend more time with their smartphone unlocked. \textit{loc.var.log} is also significantly correlated with \textit{ent.pls} and \textit{unlocked.dur}, but not with \textit{acc.mag.rmssd}. This indicates that it is visiting and spending time at multiple places during a day, rather than merely covering more expansive geographic area which may not entail stopping, that is associated with active motion. 

\subsection{Fitbit findings}\label{subsec:fitbit_findings}

A notable functionality of Fitbit is to infer and categorize the wearer's bedtime into sleep stages: awake, non-REM sleep (including ``light" and ``deep" sleep), and REM sleep. Researchers have sought to validate Fitbit's ability to accurately measure sleep stages by concurrently monitoring Fitbit wearers' sleep using polysomnographic devices and found satisfactory accuracy of differentiating asleep and awake as well as delineating REM stage sleep \cite{de2018validation,liang2018validity}. Many others used Fitbit sleep stage data to study the relations between sleep patterns and other health and performance outcomes such as asthma \cite{bian2017exploring}, cognitive ability \cite{tucker2006daytime}, and sleep disorders \cite{silva2008sleep}. However, less is known about how the amount of REM and non-REM sleep is associated with an individual's perceived sleep quality. To help answer this question, we correlate three Fitbit-measured nighttime sleep duration metrics (REM time, non-REM time, and total asleep time that is REM + non-REM) with participants' answer to a question about self-perceived quality of the previous night's sleep (``How restful was your last night's sleep?"). This question was part of the EMA survey sent to participants at 9am each day (see Section \ref{subsubsec:ema}). Cross-referencing our Fitbit and EMA data indicates that seven participants totaling 41 nights had both reliable bedtime Fitbit data and self-reported restfulness score the next morning available. These observations are retained for the correlation analysis. 

\begin{figure}[]
    \centering
    \includegraphics[width=0.9\columnwidth]{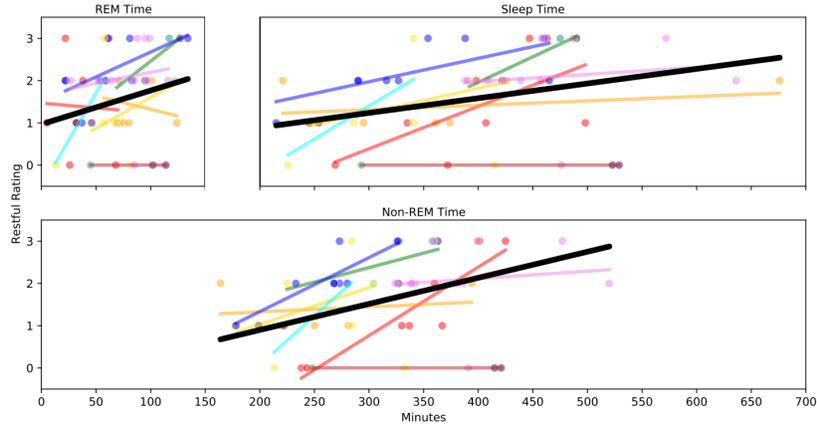}
    \caption{Relationship between participants' self-reported sleep restfulness in the morning and Fitbit-measured duration of REM, non-REM, and total sleep received the previous night. Each dot represents one night and is colored to indicate distinct participants. Colored-lines and black line represent individual and overall linear trends respectively.}
    \label{fig:fitbit-beiwe-sleep-restful}
\end{figure}

\begin{table}[]
\centering
\caption{Coefficients and significance (p-value) of the correlations between self-reported sleep quality (restfulness) and Fitbit-measured sleep metrics based on mixed-effect ordinal regression. }
\begin{tabular}{@{ }lll@{ }}
\toprule
Fitbit-measured duration (hours) & coefficient ($\beta$) & p-value\\ \midrule 
REM &  1.312 &  0.04  \\ 
Non-REM & 1.136 & 0.003  \\ 
Asleep (REM + non-REM)  &  0.535 &  0.012\\\bottomrule 
\end{tabular}
\label{tab:fitbit_sleep_outcomes_ordinal}
\end{table}

For each participant-night, the duration of Fitbit-measured REM, non-REM, and total sleep time (in minutes) were paired with the perceived restfulness self-rating the next morning (on a 0-3 ordinal scale) and plotted in three scatter plots shown in Figure \ref{fig:fitbit-beiwe-sleep-restful}. The REM time plot and the total sleep time plot are arranged abreast one another because total sleep time is always greater than REM-time and the two measures occupy two disjoint spaces on the time axis. Each dot represents one night and is color-coded to represent distinct participants. Colored linear lines are fitted specific to each participant while the thick black line represents the linear trend over all observations. Visibly there is a strong positive correlation between all three Fitbit-measured sleep duration metrics and perceived restfulness, both individually and overall. Exceptions exist in REM time plot where we observe two participants who had a non-positive correlation between REM time and perceived restfulness. To be rigorous, we further conducted ordinal regression modeling and present our results in Table \ref{tab:fitbit_sleep_outcomes_ordinal}. Since the restfulness score is an ordinal variable and multiple observations belong to the same participants, we used mixed-effect ordered logit models to model perceived restfulness rating with each of the three Fitbit-measured sleep duration metrics, with the participant-wise random effect accounted for: $logit[P(restful \leqslant i)] = \alpha_{i} - (\beta \times FitbitSleepMetric + u + \epsilon), i \in \{0,1,2\}$. Likelihood ratio tests indicated that this is superior to a fixed effect model with the corresponding Fitbit metric being the only predictor. Based on results shown in Table \ref{tab:fitbit_sleep_outcomes_ordinal}, duration of non-REM sleep appears the most significantly correlated with self-reported restfulness.

\begin{figure}[]
    \centering
    \includegraphics[width=\columnwidth]{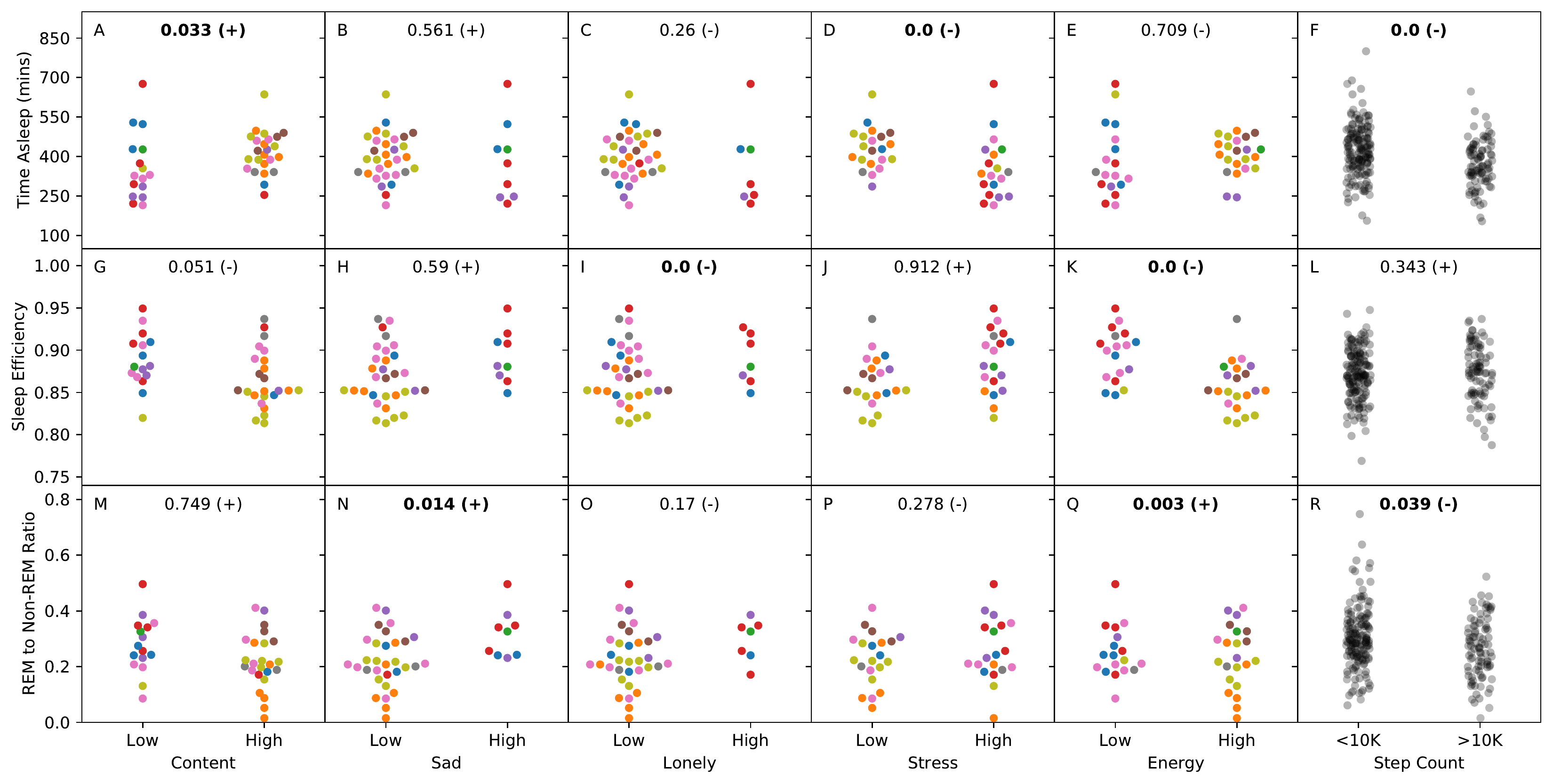}
    \caption{Differences in Fitbit-measured sleep metrics on days when participants rated aspects of their mood and energy as low ($\leqslant1$, no more than ``a little bit") or high ($\geqslant2$, at least ``quite a bit") and walked more or less than 10,000 steps. Colors in the swarm plots correspond to data from individual participants. P-values from a mixed effect models accounting for the individual random effect are included in addition to the direction of the correlation. Bold values indicate p-values less than 0.05.}
    \label{fig:fitbit_sleep_outcomes}
\end{figure}

Another group of studies have focused on the relationship between sleep quality and mental health, and found that negative mood such as anxiety, sadness, stress, and loneliness are all associated with poorer sleep quality \cite{thomsen2003rumination,akerstedt2006psychosocial,zawadzki2013rumination}. These studies typically (1) investigated between-subjects variances treating an individual participant as the unit of analysis and (2) used either questionnaires (e.g., Pittsburgh Sleep Quality Index) or in-lab sensors (e.g., polysomnography, actigraphy) to assess sleep quality. We, on the other hand, wanted to explore how an individual's mood, energy, and physical activity levels assessed on a \textit{daily} basis may influence her sleep quality on the immediately following night as measured by the more convenient and less intrusive Fitbit. Specifically, we calculated three sleep metrics based on Fitbit sleep stage data: (1) time asleep, which is the sum of both REM and non-REM time; (2) sleep efficiency, defined as the ratio of time asleep to time in bed, and (3) the REM-to-non-REM ratio. For daily mood and energy ratings, we used participants' end-of-each-day EMA self-reports of contentment, sadness, loneliness, stress, and energy levels, evaluated on a 0-3 ordinal scale (with the exception of energy level which was evaluated on a 0-4 ordinal scale); for each of the mood and energy questions, we labeled a day as ``low" if the participant's self-rating was below or equal to 1 (i.e., no more than ``a little bit") and ``high" if greater than or equal to 2 (i.e., at least ``quite a bit"). For physical activity level, we used the daily step count also captured by Fitbit, and labeled a day as ``low" if the participant registered less than or equal to 10000 steps and ``high" otherwise, which is consistent with a guideline widely used in health research \cite{wattanapisit2017evidence}. We then compared each of the three nightly Fitbit sleep metrics with each of the binary mood, energy, and activity levels of the day preceding sleep. Requirements for data availability for Fitbit sleep stage data, daily EMA data, and Fitbit step count data narrowed our sample size down to 41 observations (days) from 7 participants for the sleep-EMA pairing and 252 observations from 34 participants for the sleep-steps pairing. 

The distributions of the three Fitbit sleep metrics given high or low mood, energy, and step counts are shown in Figure \ref{fig:fitbit_sleep_outcomes}. Mood/energy and sleep metric relationships are shown as swarm plots with data points colored by participant. Step count and how it relates to sleep metrics are shown in the final column of Figure \ref{fig:fitbit_sleep_outcomes} as a pair plot. In addition to plotting, we built mixed effect models of each sleep metric using each of the mood, energy, and step count binary levels with the individual participant's random effect accounted for and show the sign and p-value of the coefficient in each corresponding cell in Figure \ref{fig:fitbit_sleep_outcomes}. Class imbalance between high and low sadness and loneliness is visibly severe (panels B, C, H, I, N, O), undermining the reliability of the corresponding results, therefore we limit our interpretation to the remaining four explanatory variables (contentment, stress, energy, and step count). Contentment and stress share a similar pattern in terms of correlation with the Fitbit sleep metrics: they are both significantly correlated with time asleep (panels A and D, albeit in opposite directions) but not with sleep efficiency or REM-to-non-REM ratio (panels G, M, J, P). A day reported by a participant as higher in contentment and lower in stress tends to end with significantly greater amount of time asleep at night, which is consistent with existing research on negative mood and sleep duration \cite{kim2007effect,vandekerckhove2011role,zhai2015sleep}. The only variable that has a significant correlation with both sleep efficiency and REM-to-non-REM is energy level during the day (panels K and Q). Days on which participants reported feeling more energetic tend to have worse sleep efficiency (e.g., longer time tossing and turning before falling asleep) but a higher REM-to-non-REM ratio. The final column of plots show that on days when participants took more than 10000 steps, their sleep time was shorter and had a lower REM-to-non-REM ratio, which seems to contradict the finding in most studies that increased exercise is correlated with improvements in sleep duration and quality \cite{dolezal2017interrelationship,kelley2017exercise}. This result suggests that steps taken on a day may not be an effective indicator of the amount of physical exercise, at least among college students. 

\subsection{Home environment findings}\label{subsec:bevo_findings}

\begin{figure}[]
    \centering
    \includegraphics[width=\columnwidth]{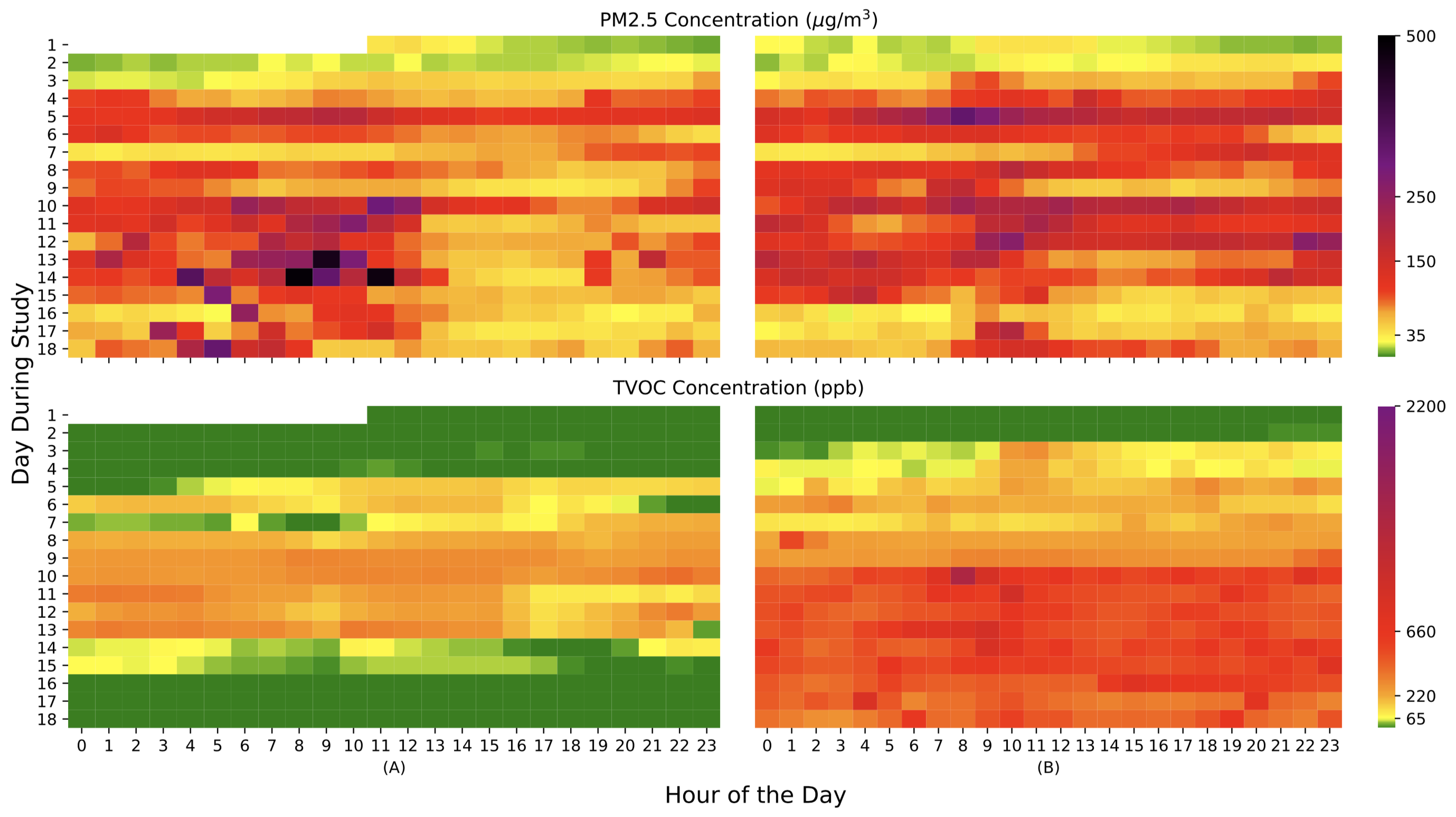}
    \caption{Hourly and daily variations of PM2.5 and TVOC concentration from three BEVO Beacons throughout the data collection period. The color of a cell indicates the environmental hazard associated with a certain level of PM2.5 and TVOC.}
    \label{fig:beacon_iaq_heatmap}
\end{figure}

We begin by plotting data from two example participants' BEVO Beacons (IDs \textit{vvlm8iyi}, \textit{19k2rlua} in Figure \ref{fig:data_reliability}) in Figure \ref{fig:beacon_iaq_heatmap} to show the hourly variation of indoor air quality throughout the study period. We chose these participants because their devices collected the most complete sets of environmental data and their responses on the Home Environment and Health survey revealed no major differences in their living situations: each participant had one roommate, did not smoke, did not have any carpeting, did not own pets, and did not regularly change their AC air filter. Figure \ref{fig:beacon_iaq_heatmap} shows the variation of PM2.5 and TVOC concentrations over different days and different hours of the day for the three participants. The colors in Figure \ref{fig:beacon_iaq_heatmap} correspond to air quality index (AQI) values for PM2.5 \cite{mintz2009technical} and approximated AQI values for TVOC. For PM2.5, AQI is a scale from 0 to 500 that indicates the severity of pollutant concentration. An AQI of 100 is consistent with recommendations from the US Environmental Protection Agency's (EPA) National Ambient Air Quality Standards (NAAQS) therefore values for AQI below 100 are generally considered healthy while values above can be a concern depending on the concentration and duration of the exposure.
 
All three participants experienced consistently high concentration of PM2.5 on days such as the fifth and tenth days of the study compared to other days. The participant in column (A) saw decreases in PM2.5 concentration every day starting after 12:00 with the exception of days 5 and 7. One plausible explanation is that the participant's AC turns on at this time and remains operating during the hottest part of the day. Weather data indicate that temperatures on day 5 were lower, on average, than most other days during the study period which could have made running the AC unnecessary. Between days 10 and 18, this participant's BEVO Beacon captures 1-3 hour events where concentrations were uncharacteristically high. These typically occur in the early-to-late morning which may indicate the participant was cooking breakfast and the effects of these events lingered for a few hours after. There are no noticeable trends for the participant in column (B), however their PM2.5 concentrations decrease starting at 12:00 on day 15 and remain suppressed for the remainder of the study.

One shared pattern between PM2.5 and TVOC measurements is an initial period of low value readings. Lower PM2.5 concentrations were grouped toward the beginning of the study period and not measured again for the remainder of the study. TVOC concentrations are also similarly low during the first few days of the study. This pattern could be explained by either or both of the following. First, sensors may have required significant time to warm-up; for example, metal oxide sensors used to measure TVOCs, like the SGP30 sensor on the BEVO Beacon, typically have a burn-in time meaning that when the sensors are initially powered on, measurements are uncharacteristically low. Second, sensors may have experienced a certain degree of \textit{drift} or \textit{fouling} that artificially increased the measured concentration during later portions of the study: sensor drift refers to the gradual increase or decrease in measured values over a long period of time due to degradation of electrical and mechanical components, while fouling occurs when sensors are exposed to dramatically high concentrations or in operation for long periods of time which allows dust and other materials to build up around the sensors inlet. The TVOC plots in columns (B) suggest that the sensor might have become over-saturated because the concentrations get progressively higher throughout the study period. These concentrations, like PM2.5 of (B), are also dramatically higher than the values measured for the participant in column (A) who experienced worse concentrations during the middle of the study period and a decrease during the final few days.

\begin{figure}[]
    \includegraphics[width=\columnwidth]{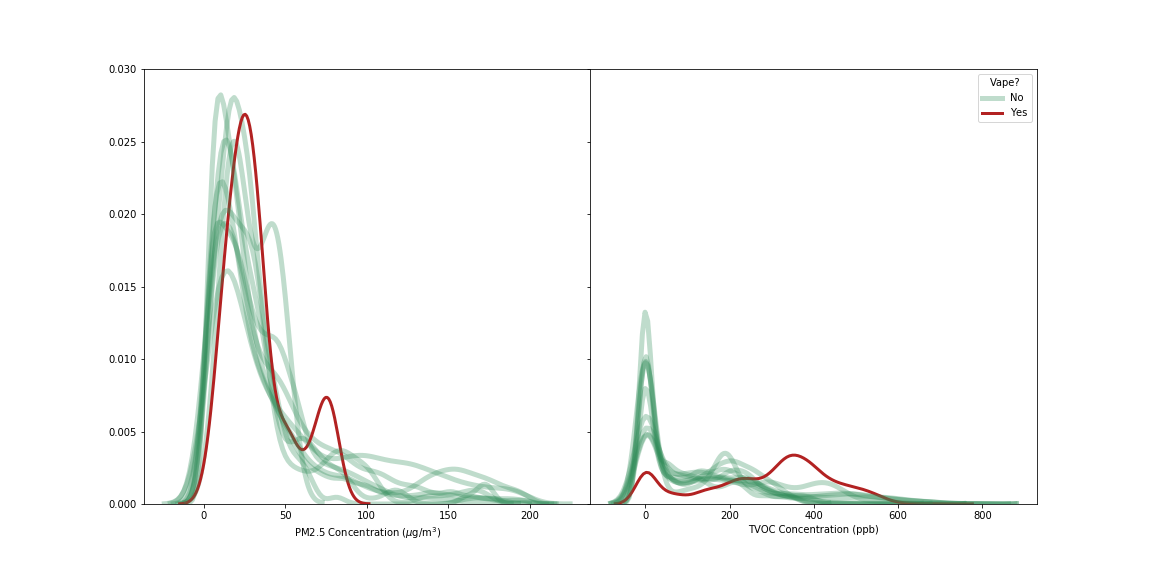}
    \caption{Comparison of distributions of PM2.5 and TVOC measurements in homes between participants who do not vape and those who do (one participant, indicated by red curve).}
    \label{fig:iaq_vape_trends}
\end{figure}

Figure \ref{fig:iaq_vape_trends} shows the distinctive distribution densities of PM2.5 and TVOC concentration values in the living environment of the vaping participant compared to the concentrations measured in nine other non-vaping participants, suggesting that our BEVO Beacon can potentially be used to detect living habits and inform behavioral health interventions. Recent studies have indeed measured increased concentrations of PM2.5 and TVOCs in rooms after vaping \cite{eaton2018cigarette,chen2018assessment}. Only one participant out of the ten who received a BEVO Beacon during the Spring 2019 deployment also reported that they vaped. This vaping participant had an elevated density of PM2.5 concentrations around 75 $\mu$g/m$^3$ but did not record values as high as other participants. Compared to other participants, the vaper also recorded higher TVOC concentration on average and especially an increased proportion of TVOC concentration between 300-500 ppb. Our results, while limited to one individual, are consistent with previous study findings that vaping increases the amount of PM and TVOCs in one's indoor environment. 

\section{Discussion}\label{sec:discussion}

We have focused on four types of human-centric data and their inter-connections in our analysis, namely EMA self-reports, smartphone sensing, wearable sensing (Fitbit), and home environment sensing (BEVO Beacon). The four data modalities serve as an illustrative subset of the space outlined by our conceptual framework (Figure \ref{fig:framework}) and are not meant to be exhaustive. For example, we collected single-time, highly participatory measures such as buccal swabs but chose not to include them in our data experiments due to the limited available sample size. Future work could very well place heavier emphasis on obtaining these data types that are usually more logistically challenging to collect and explore how key well-being outcomes revealed in those data types can be inferred by more passive data streams. Another type of data that we did not collect in the UT1000 Project is individual social media data including both the content created and the interaction patterns, which have been utilized to predict personal mental health outcomes in past studies \cite{mendu2019socialtext}. In our conceptual framework of human-centric data modality, social media data would fit in the medium range on both the temporal coverage and the spatial freedom axes, similar to the position of EMAs. 

A recurring theme in the data analysis using multi-modal human-centric data, as we encountered throughout Section \ref{sec:results}, is multiple measurements from the same participant. Typically, when researchers collect sensing or survey data from participants, multiple data points for each data type are collected from each participant, resulting in a situation where one has $N$ observations from $M$ participants and $N > M$. This is not a trivial issue because observations that belong to one participant tend to be more correlated with each other than with those from other participants, which makes the direct application of regression models onto the entirety of the $N$ observations statistically unsound \cite{bolger2013intensive}. We identify the following strategies for dealing with this technical challenge. First, if the inter-individual variation in a particular variable is of interest and the number of participants $M$ with data available is satisfactorily high, one appropriate approach is to aggregate the variable by the individual, such that each individual is mapped to only one value (e.g., mean value) of that variable. This is the approach we adopted for the analysis in Section \ref{subsec:ema_findings}. Interpretations made from models created on such aggregated data would be regarding the cross-participant patterns rather than within-participant ones. Second, if one has limited amount of data available that does not justify aggregation, an appropriate solution is to use mixed effect models to account for individual differences between participants. We used a mixed effect ordinal regression models in Section \ref{subsec:fitbit_findings}. An individual random effect could be added to both the intercept or a variable coefficient of a (generalized) linear model and the eventual selection will need to depend on model selection metrics or tests such as AIC, BIC, or a likelihood ratio test to determine whether the added effect (thus increasing variance explained) is worth the increased model complexity. In our experiments discussed in Section \ref{subsec:fitbit_findings}, specifically those presented in Table \ref{tab:fitbit_sleep_outcomes_ordinal} and Figure \ref{fig:fitbit_sleep_outcomes}, we found that adding only an individual random effect to the intercept is sufficient and the most appropriate. 

Most of the analysis done in this paper (Section \ref{sec:results}) falls in the category of \textit{correlation analysis}, where one builds statistical models to fit to all data to learn the correlations between variables, and evaluate the models using some goodness-of-fit metric (e.g., adjusted R squared, significance). Another type of tasks researchers often conduct with multi-modal human-centric data is \textit{predictive modeling}, for which the multiple measurement issue may have different and nuanced implications. In a predictive modeling task, one builds (trains) a model to fit to a subset of all data and seeks to evaluate (test) the predictive power of the model using the remaining subset with a prediction performance metric (e.g., area under ROC curve). When the multiple measurement issue is present (i.e., $N > M$), the train-test partition needs to be carefully executed based on the nature of the predictive task at hand. If the predictive modeling task focuses on detecting (especially previously unknown) individuals that belong in a certain class (e.g., clinical diagnosis), it is advisable to assign observations from different participants into the training and test set because having the same participant's (thus correlated) observations in both sets data will inflate predictive performance. However, if the predictive task aims to monitor a participant's status and raise warnings when undesirable changes are detected, then the correlation between observations from the same participant becomes a valuable source of information to capitalize on: by training models using past observations from a participant, future observations of the same individual, being correlated with her previous ones, become easier for the models to classify. This is why personalized models typically perform better than generic models for personal monitoring tasks \cite{constantinides2018personalized}. To achieve decent performance for a personal monitoring task, an initial period of model training tuning is usually necessary (as opposed to a ``cold start"); sometimes, the optimal strategy for improved prediction performance may be a hybrid model using both other participants' data and the same participant's previous data.

\section{Conclusion}

We conducted the UT1000 Project, a multi-modal data collection study using a variety of technologies and methods to monitor and understand aspects of the health, behavior, and living environments (See Figure \ref{fig:example_pid}) of a college student cohort of more than 1500 participants for 3 weeks. Some participants voluntarily monitored themselves for several more weeks after the official 3-week study period ended. The project is highly novel due to not only the large scale of participation but also the emphasis on incorporating the monitoring of personal living environment with health and behavior sensing to achieve a multi-faceted dashboard of human-centric information. With many types and sources of data at hand, we proposed a conceptual framework systematically organizing human-centric data modalities and their corresponding technologies and methods with respect to their temporal coverage and spatial freedom, which is further helpful for guiding data collection and research question formulation. Temporal coverage and spatial freedom overlap with ecological validity and are constrained by the unobtrusiveness of the method. Hurdles to unobtrusiveness include the size, weight, and power need of a device, requirement for human attention and maintenance, and many potential others. A general direction of evolution for human-centric design and technology is to become more portable, convenient, and user-friendly thus affording higher and higher unobtrusiveness and ecological validity for its capacity of understanding and assisting humans, until it is truly ``woven into the fabric of everyday life" \cite{weiser1999computer}.

We were able to collect from a large participant cohort satisfactorily complete multi-modal data in terms of both data continuity and participant compliance (see Figure \ref{fig:data_reliability} and discussions in Section \ref{subsec:data_completeness}). Our findings with EMA data point to differential emotional experience associated with the places in which one spends time and the people with whom one spends time. Certain aspects of emotions are more interlinked than others; for example, a person's sadness is especially connected with feelings of loneliness but less so with contentment and stress. Our smartphone data confirms the benefit of a broadly construed ``activeness" in life, whereby individuals that move more and spend time at more different places generally enjoy a more healthy mind. Fitbit findings support the utility of Fitbit's sleep measuring capability and reveal several significant correlations between Fitbit sleep metrics and self-reported sleep restfulness and mood outcomes. Lastly, the environmental sensors in our BEVO Beacons are capable of detecting abnormalities in an individual's indoor living environment, some of which are useful for detecting daily activities and lifestyle choices. 

Several limitations exist in our study, which we would like to address in future work. First, we would like to monitor participants for a longer period of time than three weeks so that we are be able to observe more reliable patterns of behavioral variation and build more accurate personalized predictive models. Second, there is a sharp imbalance between the availability of smartphone and EMA data and the availability of Fitbit and BEVO Beacon data due to our incentive structure and hardware availability. As a result, analyses involving Fitbit and BEVO Beacon data encountered the problem of small sample size and our results are of a preliminary and exploratory nature rather than final verdicts. We are preparing for a third deployment of the UT1000 Project that directly addresses these limitations by recruiting (potentially fewer) participants who will commit to longer study periods as well as increasing the number of Fitbits and BEVO Beacons distributed. We anticipate the collection, integration, and mining of diverse modalities of human-centric data from different technologies and methods and of various degrees of temporal coverage and spatial freedom to be key to the development of a new generation of digital solutions for personal well-being enhancement. 

\section*{Funding}
This work was supported by Whole Communities—Whole Health, a research grand challenge at the University of Texas at Austin, and National Science Foundation Award SES-1758835.

\bibliographystyle{plain}
\bibliography{main}

\newpage

\section*{Appendix A: HEH Survey Questions}

\begin{table}[h]
\scriptsize
\centering
\begin{tabular}{p{0.6\linewidth}p{0.1\linewidth}p{0.1\linewidth}p{0.1\linewidth}}
\toprule
Question text & Variable Type & Answer Options & Example \\ \midrule
Which best describes your current living situation: & Str & Apartment, Dormitory, Stand-alone House & Apartment \\
How many roommates?  & Int & Enter Value & 2 \\ 
How many housemates?  & Int & Enter Value & 2 \\
How many of your housemates and roommates are female?  & Int & Enter Value & 2 \\
How many of your housemates and roommates are male?  & Int & Enter Value & 0 \\
How many of your housemates and roommates are nonbinary?   & Int & Enter Value & 1 \\
Does anyone smoke cigarettes in the house?  & Str & Yes, No & No \\ 
Does anyone vape in the house? & Str & Yes, No & No \\ 
Do you have any pets?  & Str & Cat, Dog, Other, N/A & Dog \\ 
How many dogs? & Int & Enter Value & 0 \\ 
How many cats? & Int & Enter Value & 0 \\ 
How many `other' pets? & Int & Enter Value & 0 \\ 
What floor is your living space on (count ground floor as 0 and count up from there)? & Int & Enter Value & 5 \\ 
Does the door to your apartment connect to the: & Str & indoor corridor (not open to the outside), indoor corridor (open to the outside), outdoors & Housemates \\ 
Does your living space have central air conditioning and heating?  & Str & Yes, No & No \\ 
If yes, do you have control over changing the air conditioning filter?  & Str & Yes, No & No \\ 
If yes, how often do you change it (in months)? & Str & Enter Response & Varies \\ 
If yes, how do you choose the filter (price, quality)? & Str & Enter Response & Varies \\ 
If yes, which filter do you choose (brand and specific product)? & Str & Enter Response & Varies \\ 
\end{tabular}
\end{table}

\begin{table}[]
\scriptsize
\centering
\begin{tabular}{p{0.6\linewidth}p{0.1\linewidth}p{0.1\linewidth}p{0.1\linewidth}}
If no, how often does your apartment manager change it?  & Str & Enter Response & Varies \\ 
Did you receive a swab kit as a participant in the  extended study? & Str & Yes, No & No \\ 
If yes, did you swab your: & Str & AC vent, AC filter & AC Vent \\
Do you have control over the temperature of the home? & Str & Yes, No & No \\ 
If yes, how often per week do you typically set/change it? & Int & Enter Value & 3 \\
If yes, are you the only one who uses the thermostat? & Str & Yes, No & No \\ 
If yes, what temperature is normally chosen in winter (degrees F)? & Int & Enter Value & 77 \\ 
If yes, what temperature is normally chosen in summer (degrees F)? & Int & Enter Value & 79 \\
Do you open your windows to ventilate your home? & Str & Yes, No & No \\ 
If yes, how many times per week? & Int & Enter Value & 3 \\ 
Does your living space have water damage? & Str & Yes, No & No \\ 
If yes, indicate what space has water damage (room type, surface type): & Str & Enter Response & Varies \\
Does your home have moldy odor when you enter the space? & Str & Yes, No & No \\ 
Has anyone commented on bad odors when entering your home? & Str & Yes, No & No \\ 
Do you use `air cleaner' devices? & Str & Yes, No & No \\ 
Does your house have carpet? & Str & Yes, No & No \\ 
If yes, about what \% of the total area is carpeted? & Int & Enter Value & 50 \\
Does your house have: & Str & hardwood, tile floors & hardwood \\ 
If yes, about what \% of the total area is hardwood? & Int & Enter Value & 25 \\
If yes, about what \% of the total area is tile floor? & Int & Enter Value & 25 \\
Do you cook at home?  & Str & Yes, No & No \\ 
Do you turn on the kitchen exhaust fan? & Str & Yes, No & No \\ 
In the past three weeks, have you had the flu? & Str & Yes, No & No \\ 
Have you gotten your flu shot this year? & Str & Yes, No & No \\ 
If no, are you planning on getting the flu shot? & Str & Yes, No & No \\ 
In the past three weeks, have you caught a cold? & Str & Yes, No & No \\ 
In the past three weeks, have you suffered from allergies? & Str & Yes, No & No \\ 
In the past three weeks, have you suffered from a gastrointestinal illness? & Str & Yes, No & No \\ 
In the past three weeks, have you taken antibiotics? & Str & Yes, No & No \\ 
Do you suffer from asthma (doctor-diagnosed)? & Str & Yes, No & No \\ 
Did you receive a silicon band as a participant in the  extended study? & Str & Yes, No & No \\ 
If yes, did you consistently wear it as indicated? & Str & Yes, No & No \\ 
Do you use perfumes, cologne, or body lotions? & Str & Yes, No & No \\ 
If yes, how often per day do you apply these products? & Int & Enter Value & 2 \\
How often do you wash your hands per day? & Int & Enter Value & 10 \\
How often do you take a shower or bathe per week? & Int & Enter Value & 10 \\
Do you use electric scooters? & Str & Yes, No & No \\ 
If yes, how many times a day? & Int & Enter Value & 1 \\ 
If yes, how many times per week? & Int & Enter Value & 1 \\
\bottomrule
\end{tabular}
\end{table}

\newpage

\section*{Appendix B: EMA Questions}

\begin{table}[h]
\scriptsize
\centering
\begin{tabular}{p{0.08\linewidth}p{0.08\linewidth}p{0.2\linewidth}p{0.5\linewidth}p{0.1\linewidth}}
\toprule
Survey Distribution & Question Type & Question Text & Answer Options & Example \\ \midrule
Morning only & Radio Button & How many hours did you sleep LAST NIGHT? & 0 hours; did not sleep; 1-2  hours; 2-3 hours; 3-4 hours; 4-5 hours; 5-6 hours; 6-7 hours; 7-8 hours; 8-9 hours; 9-10 hours; 10-11 hours; 11-12 hours; more than 12 hours & 8-9 hours \\
Morning only & Radio Button & How restful was your sleep? & Not at all restful; Slightly restful; Somewhat restful; Very restful & Somewhat restful \\
Morning only & Radio Button & How refreshed did you feel after your sleep? & Not at all refreshed; Slightly refreshed; Somewhat refreshed; Very refreshed & Somewhat refreshed \\
All & Checkbox & please describe your behavior during the PAST FIFTEEN MINUTES... I spent MOST of my time in the following place: & Bar; Party; Cafe; Restaurant; Campus; Fraternity; Sorority House; Gym; Home (dorm; apartment); Library; Religious facility; Store / Mall; Work; Vehicle; Friend; None of the above; other & other \\
All & Checkbox & please describe your behavior during the PAST FIFTEEN MINUTES...I spent MOST of my time with the following people: & Classmates; students; Co-workers; Family; Friends; No one; alone; Roommates; Significant other; Strangers; Other & Friends \\
All & Checkbox & please describe your behavior during the PAST FIFTEEN MINUTES...I spent MOST of my time & Attending classes; meetings; Browsing the Internet; using social media; Commuting; traveling; Doing household chores; running errands; Eating; drinking; Exercising; physical activity; sports; Resting; napping; doing nothing; Studying; reading; preparing for an exam; Talking; texting; socializing; Watching TV; movies; Working at job; None of the above; Other & napping \\
All & Checkbox & please describe your behavior during the PAST FIFTEEN MINUTES I spent time interacting with others by: & Talking in person; Talking on the phone; Chatting on Whatsapp or other chat app; Chatting on a dating app; Emailing; Video-chatting; Interacting on Facebook; Interacting on Instagram; Interacting on Snapchat; Interacting on Twitter; Other form of social interaction; Not applicable; was not interacting with anyone & Talking in person \\
All & Radio Button & I am feeling CONTENT: & Not at all; A little bit; Quite a bit; Very much & Very much \\
All & Radio Button & I am feeling STRESSED: & Not at all; A little bit; Quite a bit; Very much & Quite a bit \\
All & Radio Button & I am feeling LONELY: & Not at all; A little bit; Quite a bit; Very much & A little bit \\
All & Radio Button & I am feeling SAD: & Not at all; A little bit; Quite a bit; Very much & Not at all \\
All & Radio Button & My ENERGY LEVEL is: & Low energy; Somewhat low energy; Neutral; Somewhat high energy; High energy & High energy \\
\bottomrule
\end{tabular}
\end{table}

\newpage

\section*{Appendix C: Beiwe Sensing Parameters}

A comprehensive description of the passive data the Beiwe platform collects can be found on the developer's Wiki platform: https://github.com/onnela-lab/beiwe/wiki/Passive-Data. A short description of the data collected and used in this study is summarized below.

\begin{table}[h]
\scriptsize
\centering
\begin{tabular}{p{0.14\linewidth}p{0.42\linewidth}p{0.1\linewidth}p{0.3\linewidth}}
\toprule
Data Label & Short Description & Operating Systems & Variables Collected \\ \midrule 
Accelerometer & Indication of participant movement & iOS, Android & timestamp, accuracy, x, y, z \\
GPS & Phone's location & iOS, Android & timestamp, latitude, longitude, altitude, accuracy \\
Power State & Phone screen, charging, or percentage of battery & iOS, Android & timestamp, event \\
Bluetooth & Records hashed MAC addresses of nearby devices & Android & timestamp, hashed ID\\
Reachability & Phone is connected to WiFi, cellular network, in airplane mode, or no service & iOS & timestamp, event \\
\bottomrule
\end{tabular}
\label{tab:beiwe_passive_data}
\end{table}

\end{document}